\documentclass{aa}  

\usepackage{subfigure}
\usepackage{graphicx}
\usepackage{longtable}
\usepackage{units}
\usepackage{txfonts}
\usepackage{color}

\begin{document}

   \title{Effect of dust grain porosity\\on the appearance of protoplanetary disks}


   \author{F. Kirchschlager
          \and
           S. Wolf
          }

   \institute{Institut f\"ur Theoretische Physik und Astrophysik, Christian-Albrechts-Universit\"at zu Kiel,
              Leibnizstra\ss e 15, 24098 Kiel, Germany\\
              \email{kirchschlager@astrophysik.uni-kiel.de}
             }

   \date{Received 3 December 2013; accepted 9 July 2014}

 
  \abstract
{We theoretically analyze protoplanetary disks consisting of porous dust grains. In the analysis of observations of protoplanetary disks the dust phase is often assumed to consist of spherical grains, allowing one to apply the Mie scattering formalism. However, in reality, the shape of dust grains is expected to deviate strongly from that of a sphere.}
{We investigate the influence of porous dust grains on the temperature distribution and observable appearance of protoplanetary disks for dust grain porosities of up to $\unit[60]{\%}$.}
{We performed radiative transfer modeling to simulate the temperature distribution, spectral energy distribution, and spatially resolved intensity and polarization maps. The optical properties of porous grains were calculated using the method of discrete dipole approximation.}
{Grain porosity has a strong influence on the spectral energy distribution and scattered-light maps of protoplanetary disks. The flux in the optical wavelength range is for porous grains higher than for compact, spherical grains. The profile of the silicate peak at $\sim\unit[9.7]{\mu m}$ strongly depends on the degree of grain porosity. The temperature distribution shows significant changes in the direction perpendicular to the midplane. Moreover, simulated polarization maps reveal an increase of the polarization degree by a factor of about four when porous grains are considered, regardless of the disk inclination. The polarization direction is reversed in selected disk regions, depending on the wavelength, grain porosity, and disk inclination. We discuss several possible explanations of this effect and find that multiple scattering explains the effect best.}
{Porosity influences the  observable appearance of protoplanetary disks. In particular, the polarization reversal shows a dependence on grain porosity. The physical conditions within the disk are altered by  porosity, which might have an effect on the processes of grain growth and disk evolution. }
 

   \maketitle

\section{Introduction}
The observable appearance of circumstellar disks in intensity and polarized light strongly depends on the optical properties of the embedded dust particles (e.g.,  \citealt{McCaughrean2000}; \citealt{Watson2007}). An in-depth analysis of observational data of circumstellar disks requires appropriate models for the composition, shape, and structure of the dust particles (e.g., \citealt{Min2012}). Optical properties calculated using the Mie theory (\citealt{Mie}), which assumes homogeneous, compact, spherical dust grains, often do not properly fit continuum observations. The spherical symmetry causes resonances at certain scattering angles and in the spectral energy distribution (SED), which are not observed in real disks (\citealt{Min2007}). One way to overcome this problem is to consider inhomogeneous or nonspherical dust grains (e.g.,  \citealt{Hony2002}; \citealt{Bouwman2003}; \citealt{Moreno2003}; \citealt{Min2003};  \citealt{Chiar2006}).\\
Porous dust grains do reflect these properties. Their existence is motivated by the evolution of dust in circumstellar disks. It is commonly assumed that the grains in the disk initially follow a size distribution similar to that found in the interstellar medium (ISM, \citealt{Mathis1977}). With the process of dust aggregation the particles coagulate and grow to porous, fractal aggregates (e.g., \citealt{Blum2000}; \citealt{BlumWurm2008}; \citealt{Ormel2008}). Simulations of aggregation processes even show hints for an evolution of porosity in the process of dust growing from small, compact particles via large, fractal aggregates to large, compact grains (\citealt{Ormel2007}; \citealt{Wada2007, Wada2008}). On the other hand, dust collisions in high velocity ranges cause fragmentation and deformation, resulting in arbitrarily formed particles (e.g., \citealt{Dominik1997}; \citealt{Blum2000}).\\
In addition to the internal structure of grains which may be porous, for example, the shape of the particles has a strong influence on the optical properties as well. Modified absorption and extinction properties result in changes of the corresponding spectra (e.g., \citealt{BohrenHuffman83}; \citealt{Fabian2001}; \citealt{Voshchinnikov2005, Voshchinnikov2006}). In particular, the strength, width, location, and shape of the wings of the silicate features are affected by the particle shape, but also by the dust composition and size distribution of the particles (\citealt{Voshchinnikov2008}). Nevertheless, nonspherical particles are required in the ISM to explain polarization effects by the grains that are aligned to a certain degree (e.g., \citealt{Hildebrand1995}). The material in circumstellar disks originates from the ISM. Since dust evolution has not yet proceeded sufficiently in young disks, the nonspherical grains also remain in the disks (\citealt{Weidenschilling1993}).\\
Owing to the presence of nonspherical and porous particles and because the dust dominates the appearance of disks (in the continuum), it is necessary to investigate the influence of porosity. In addition, grain porosity and thus an altered absorption and scattering behavior of the dust grains might change the physical conditions within the disk and affect the disk evolution.\\
Many general examples of studies of porous particles and their scattering and absorption features exist (e.g., \citealt{Henning1993}; \citealt{Semenov2003}; \citealt{Voshchinnikov2007}). However, their influence on the physical conditions and appearance of optically thick circumstellar disks has hardly been investigated. \cite{Siebenmorgen2012} found a shift of the thermal reemission to longer wavelengths, whereas the millimeter slope is not affected if porous grains are considered. Furthermore, the silicate feature at $\sim\unit[9.7]{\mu \textrm{m}}$ is increased and slightly shifted. \cite{Min2012} examined protoplanetary disks in polarimetric images and concluded that significant differences between the appearance of disks composed of porous particles or compact spheres do exist. The influence of porosity on the $\unit[9.7]{\mu \textrm{m}}$ and $\sim\unit[18]{\mu \textrm{m}}$ features of silicate was studied by \cite{Vaidya2011}. They found an increase of the flux ratio of the $\sim\unit[18]{\mu \textrm{m}}$ to $\unit[9.7]{\mu \textrm{m}}$ peaks with increasing porosity, while the influence on the peak position and width was minor. Altogether, the studies show the existing influence of porosity, but a broader survey of various observational properties including polarization maps of various circumstellar disks has not yet been conducted.\\
The goal of this study is to investigate the influence of grain porosity on selected aspects of the appearance of circumstellar disks. The physics of scattering and absorption processes in an optically thick disk are performed using radiative transfer simulations. This article is organized as follows: We introduce the simulation software and give a description of the chosen disk and dust model in Sect. \ref{Section2}. In Sect. \ref{Section3} we present the results for the radiative transfer simulations of disks consisting of porous grains or compact spheres. In addition to temperature distributions (Sect. \ref{Tempdistribution}), spectral energy distributions,  and  intensity maps (Sects. \ref{Scatteringradiation}, \ref{DustReemission}), this section focuses on polarization maps (Sects. \ref{PolarizationMaps}--\ref{ImplicationsPolreversal}). 
We give a summary of our results in Sect. \ref{Section4}.

 
\section{Method and model}
\label{Section2}
In this section we describe the simulation software and the model setup.


\subsection{Radiative transfer}
We used the program \texttt{MC3D} for our radiative transfer simulations, which is based on the Monte Carlo method (\citealt{Wolf1999}; \citealt{Wolf2003}). The applied temperature-correction technique is described by \cite{BjorkmanWood2001}, the absorption concept by \cite{Lucy1999}, and the scattering scheme by \cite{CashwellEverett1959}. The radiation field is described by the Stokes formalism ($I$, $Q$, $U$, $V$; \citealt{Stokes1852}; \citealt{Chandrasekhar1946}). The linearly polarized intensity is defined by $I_{\textrm{p}}=(Q^2+U^2)^{1/2}$ and the degree of linear polarization $P=I_{\textrm{p}}/I$.\\
Temperature distributions, spectral energy distributions, and spatially resolved intensity and polarization maps were simulated for a circumstellar disk at a distance of $\unit[140]{pc}$, motivated by the typical distance to nearby star-forming regions (e.g., \citealt{Elias1978}).


\subsection{Disk model}
\label{diskmodel}
We applied a parameterized disk model in which the density depends on the radial distance from the central star and the distance from the disk midplane 
(\citealt{ShakuraSunyaev1973}),
\begin{equation}
 \rho(R,z)=\rho_0 \left(\frac{R}{r_0}\right)^{-\alpha} \exp{\left(-\frac{1}{2}\left[\frac{z}{h(R)}\right]^2\right)} .\label{Shakura}                                                                                                                                                                                                                   \end{equation}
Here, $R$ is the radial coordinate in the disk midplane, and $z$ the distance from the disk midplane. The density $\rho_0$ is used as a scaling factor for the total disk mass. The scale height 
\begin{equation}
h(R)=h_0\left(\frac{R}{r_0}\right)^{\beta}
\end{equation}                                                                                                                                                                                                               
increases with radius. The geometrical parameters $\alpha$ and $\beta$ characterize the radial density profile and the disk flaring. These parameters have fixed values of $21/8$ and $9/8$, respectively. The quantities $h_0$ and $r_0$ are fixed to $\unit[10]{AU}$ and $\unit[100]{AU}$. The radial extent of the disk is limited by an inner radius $R_\textrm{in}= \unit[1]{AU}$ and an outer radius $R_\textrm{out}= \unit[100]{AU}$. For the gas-to-dust-mass ratio we assumed the canonical value of 100:1 (e.g., \citealt{Hildebrand1983}).\\
This density distribution was derived originally for a vertically isothermal, hydrostatic, non-self-gravitating disk, but its applicability for circumstellar disks was justified by \cite{DAlessio1998, DAlessio1999, DAlessio2001}. Furthermore, it has been applied successfully in previous studies, such as the Butterfly-Star IRAS 04302+2247 (\citealt{Wolf2003Butterfly}; \citealt{Graefe2013}), CB 26 (\citealt{Sauter2009}), HH30 (\citealt{Burrows1996}; \citealt{Wood1998, Wood2002}; \citealt{Cotera2001}; \citealt{Pety2006}; \citealt{Madlener2012}), HK Tauri (\citealt{Stapelfeldt1998}), HV Tauri C (\citealt{Stapelfeldt2003}), IM Lupi (\citealt{Pinte2008}),  and IRAS 4158+2805 (\citealt{Glauser2008}).


\subsection{Dust model}
\label{Dustmodel}
\begin{figure*}[bhtp!]
	\centering
\includegraphics[width=0.85\linewidth]{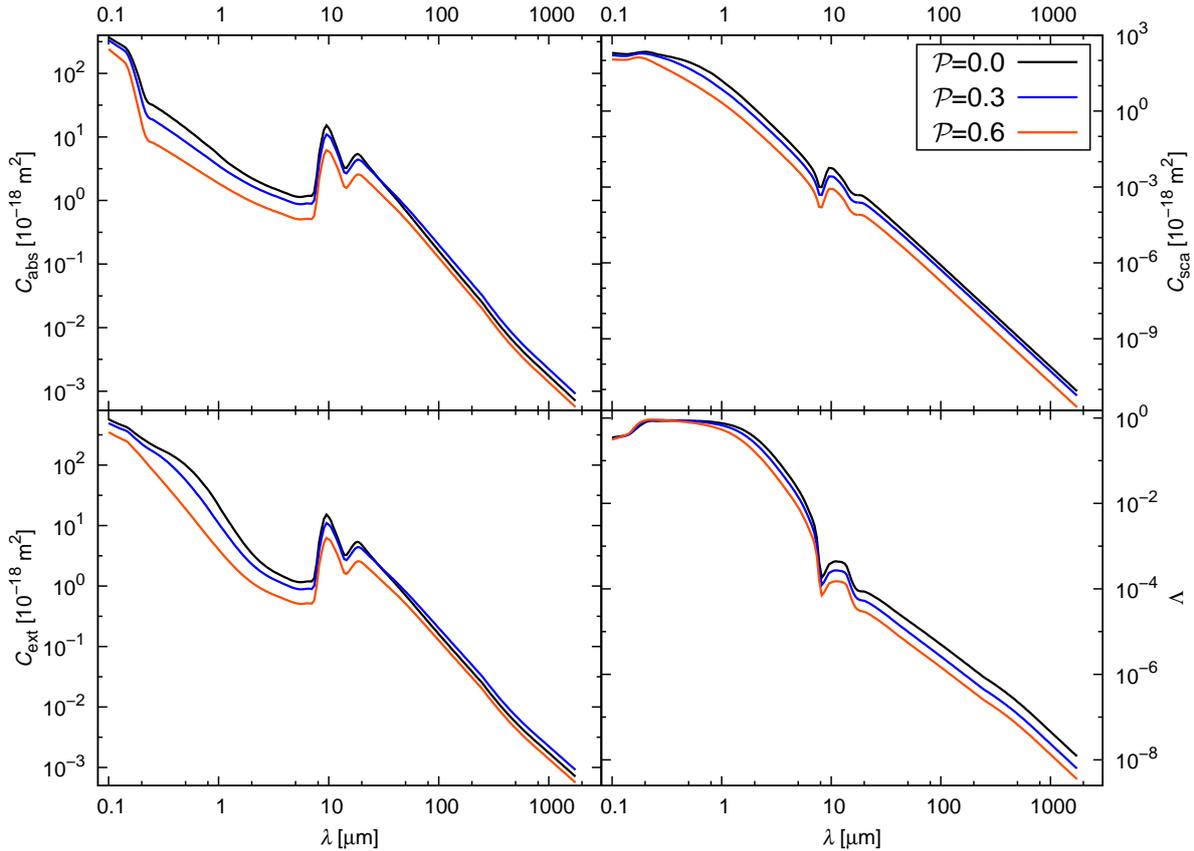}
	\caption{Absorption, scattering, and extinction cross-sections ($C_{\textrm{abs}}$, \textit{top left}; $C_{\textrm{sca}}$, \textit{top right}; $C_{\textrm{ext}}$, \textit{bottom left}), and albedo $\Lambda$ (\textit{bottom right}) of porous dust grains. The quantities are averaged based on the distribution of the particle radii, which follows a power law of  $\mathrm{d}n\propto a^{-3.5}\,\mathrm{d}a$ with $a\in \left[5,250\right] \mathrm{nm}$. The material is astronomical silicate.}
	\label{opticalproperties}
\end{figure*}
\begin{figure}[t!]
\hspace*{-0.5cm}\includegraphics[height=1.1\linewidth, angle=-90]{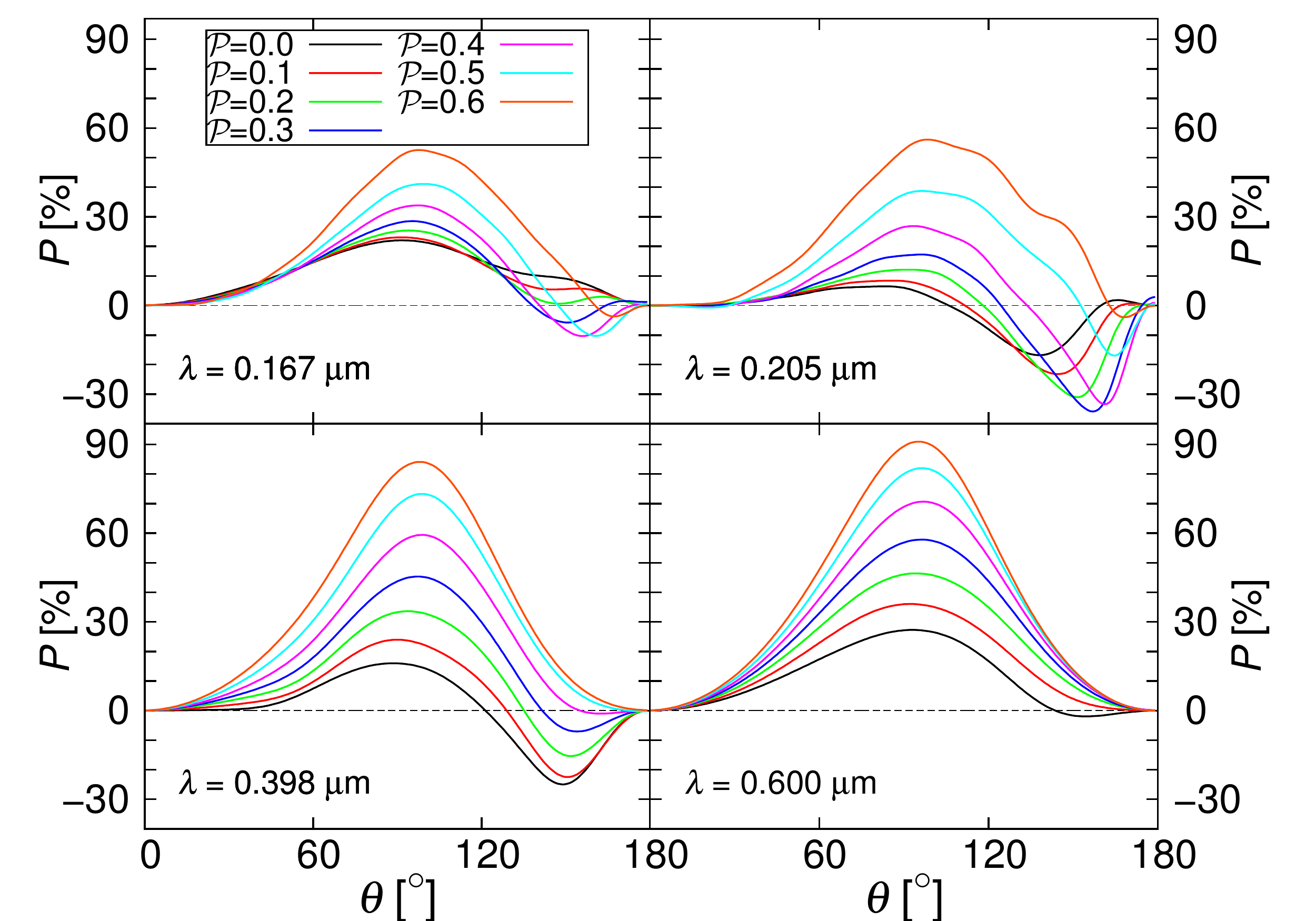}
\caption{Degree of the linear polarization $P$ as a function of the scattering angle $\theta$, the grain porosity $\mathcal{P}$, and the wavelength $\lambda$.}
\label{Polarization_lambda}
\end{figure}

The dust grains in our model are initially compact spheres from which material is removed locally, creating voids and thus a porous density distribution inside a spherical volume. We denote the radius $a$ of a particle as the smallest radius of a sphere that encases the grain. The porosity is defined as the ratio of the volume of the vacuum and the total volume of the grain,
\begin{align}
 \mathcal{P}=V_{\textup{vacuum}}/V_{\textup{total}}=1-V_{\textup{solid}}/V_{\textup{total}},\hspace*{0.5cm}\mathcal{P}\in \left[0,1\right].
\end{align}
For more information about the chosen dust model for porous grains we refer to \cite{Kirchschlager2013}. Grains with porosities of up to $\unit[60]{\%}$ were investigated. We used astronomical silicate (\citealt{Draine2003a, Draine2003b}) with a bulk density of $\rho=\unit[3.5]{g\,cm^{-3}}$ (\citealt{Draine2003}). The optical properties of the porous dust grains were calculated using the simulation software \texttt{DDSCAT} (\citealt{Draine1994, Draine2012}). It is based on the theory of discrete dipole approximation (DDA, \citealt{PurcellPenny1973}), where a continuous particle is approximated by the corresponding spatial distribution of a finite number of discrete polarizable points on a cubic lattice.\\
The dust grains follow the  size distribution $\mathrm{d}n\propto a^{-q}\,\mathrm{d}a$ with $q=3.5$ (\citealt{Dohnanyi1969}) and $a\in \left[5,250\right] \mathrm{nm}$ (MRN distribution, \citealt{Mathis1977}). A spatial variation of the grain size distribution with respect to separate processes like dust settling is not considered. To simulate the radiative transfer in the dust disk we applied the approximation by \cite{Wolf2003b}: The optical properties $\mathcal{O}(a)$ (e.g., efficiency cross-sections) of each single grain with radius $a$ are replaced by the weighted mean optical properties
\begin{equation}
\mathcal{O}_{\textrm{mean}}=\frac{\sum \limits_{a}\mathcal{O}(a)\,a^{1-q}}{\sum \limits_{a} a^{1-q}},\label{weightedmean}
\end{equation}
where the sum runs over all considered grain radii. This substantially reduces the computational time and memory effort. The resulting weighted absorption cross-sections $C_{\textrm{abs}}$, scattering cross-sections $C_{\textrm{sca}}$, extinction cross-sections $C_{\textrm{ext}}$, and albedo $\Lambda$ of porous dust grains are displayed in Fig. \ref{opticalproperties}.\\
In Fig. \ref{Polarization_lambda} the polarization $P=-S_{12}/S_{11}$ is plotted as a function of the scattering angle $\theta$, which corresponds to a single scattering of the unpolarized (stellar) light. The quantities $S_{11}$ and $S_{12}$ are the Müller matrix elements, which are a function of the wavelength and the scattering properties of the grain. The scattering angle of the highest polarization is around $90^\circ$ in most cases. Only for more compact grains at $\lambda=\unit[0.205]{\mu m}$ does the polarization maximum occur at smaller scattering angles. Except for the large scattering angles at the two lower wavelengths, the polarization increases with increasing porosity. Thus, increasing the porosity has a similar effect as decreasing the grain radius of compact spheres (\citealt{Murakawa2010}). At large scattering angles, the polarization becomes negative, which corresponds to a $90^\circ$ rotation of the polarization direction. This effect, called polarization reversal, is investigated more extensively in Sects. \ref{Sect35}--\ref{ImplicationsPolreversal}.

\subsection{Central star}
\label{Centralstar}
The circumstellar disk is heated primarily by the central star, which is assumed to be a black-body radiator with a temperature $T_*=\unit[6000]{K}$ and a radius $R_*=\unit[1.0]{{R}_\odot}$. Additional heating processes by viscosity, accretion, or cosmic radiation are not considered.

\begin{figure*}[hbtp!]
	\centering
\hspace*{-0.9cm}\includegraphics[height=0.56\linewidth, angle=-90]{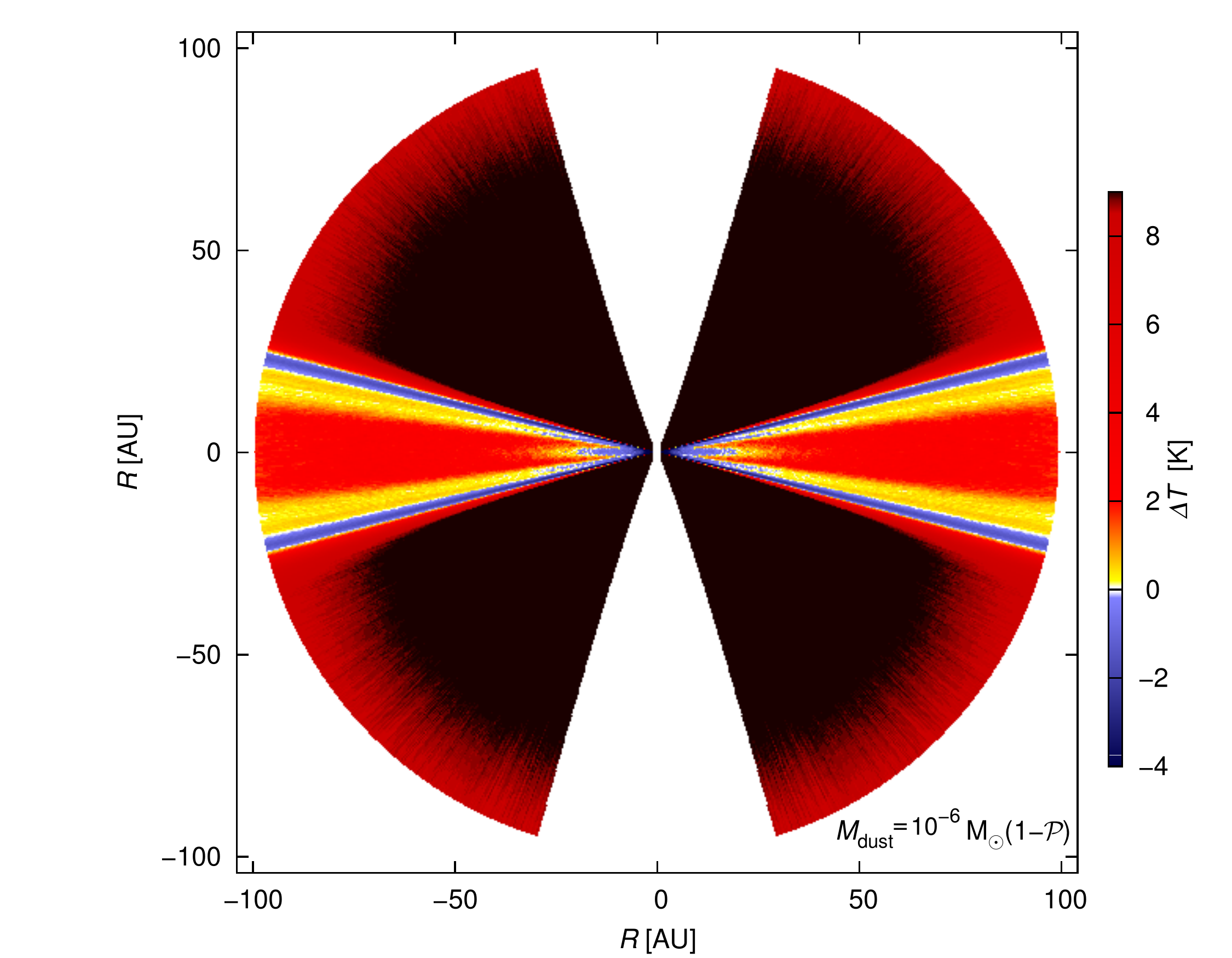}\hspace*{-1.0cm}
                \includegraphics[height=0.56\linewidth, angle=-90]{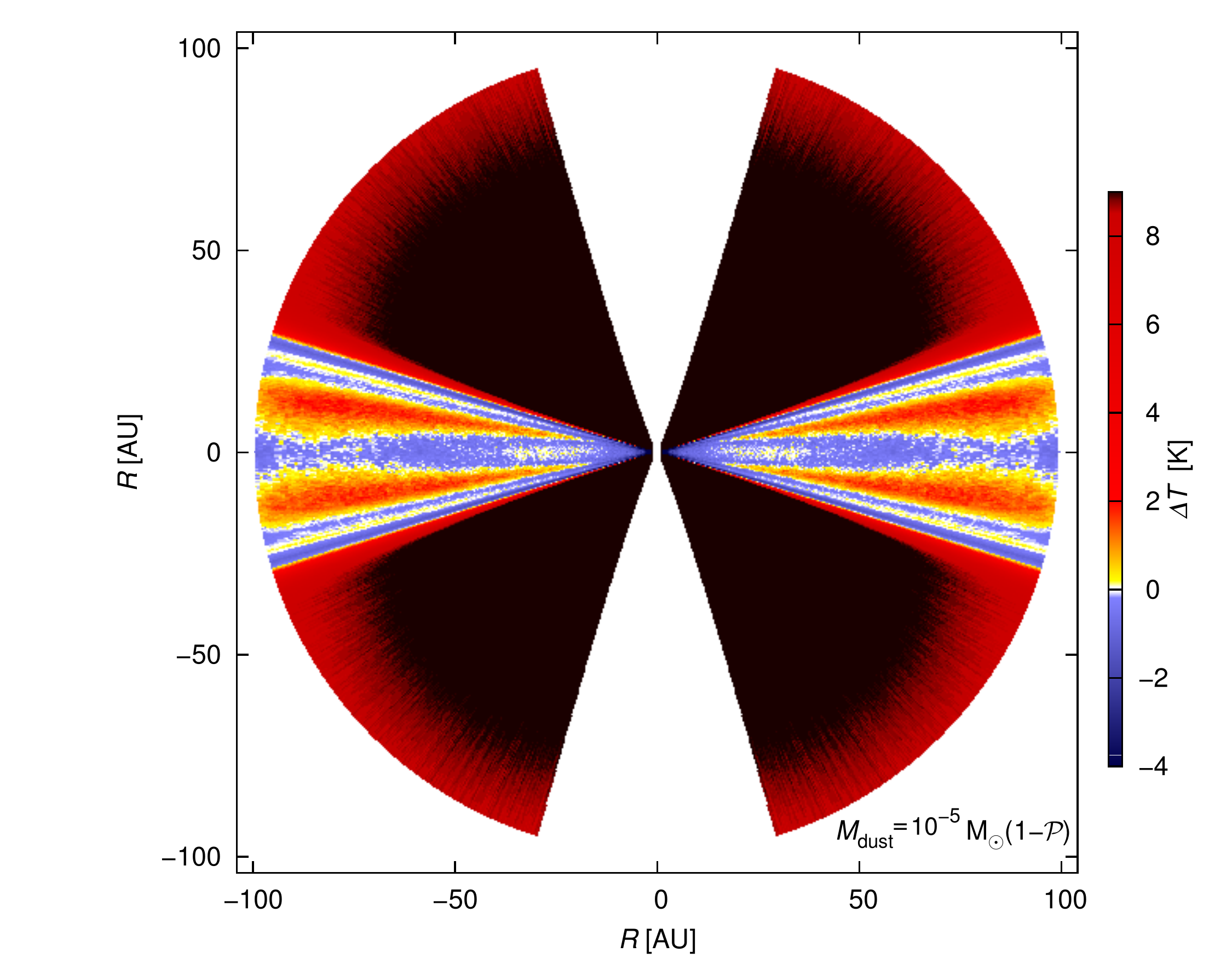}
\hspace*{-0.9cm}\includegraphics[height=0.56\linewidth, angle=-90]{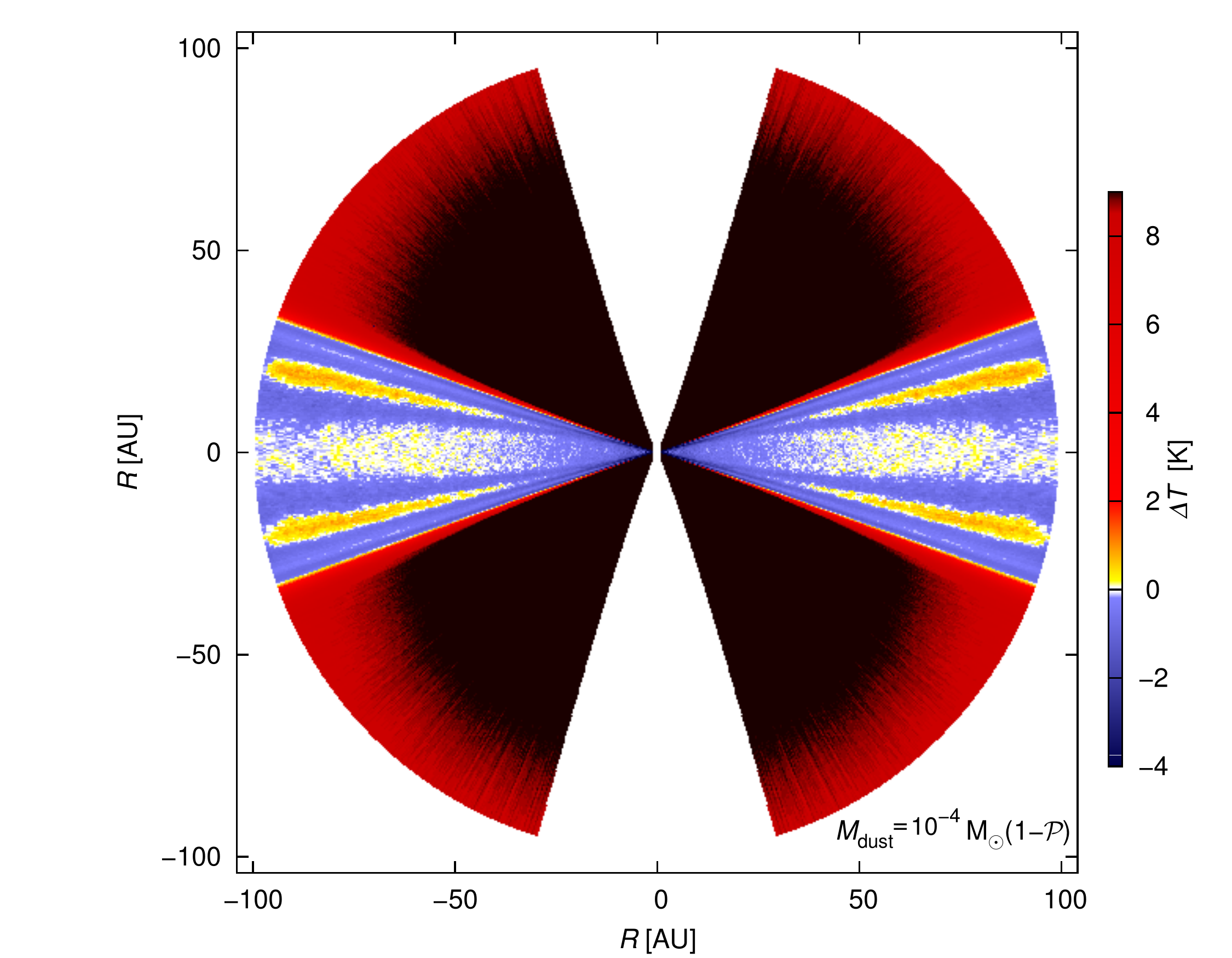}\hspace*{-1.0cm}
                \includegraphics[height=0.56\linewidth, angle=-90]{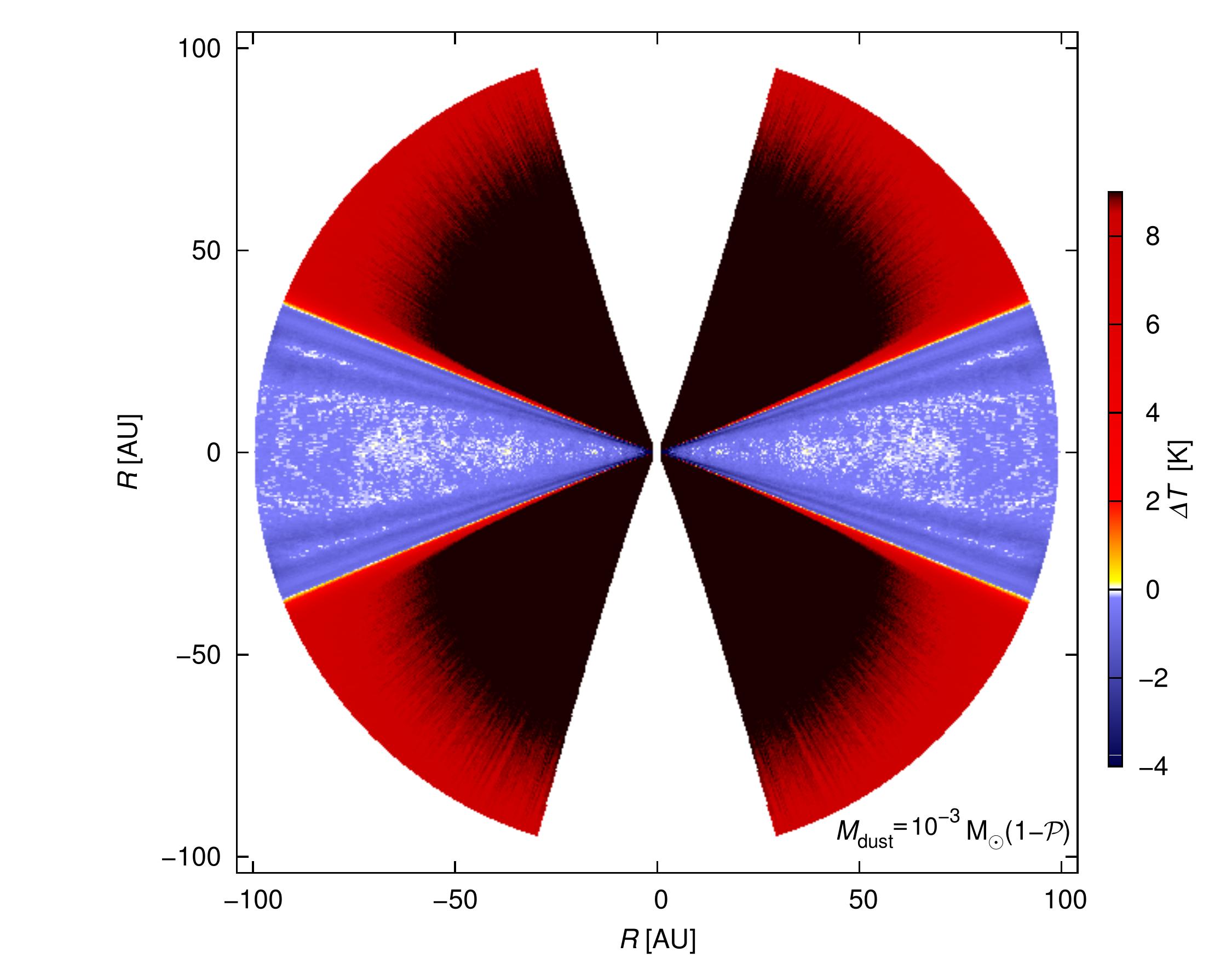}
	\caption{Difference maps of the dust temperature distributions in the plane perpendicular to the midplane. The absolute difference indicates the temperature deviations of the disks of compact spheres and porous grains, $\Delta T= T_{\mathcal{P}=0.0}-T_{\mathcal{P}=0.6}$, for  $M_{\textrm{dust}}/(\textrm{M}_\odot\, (1-\mathcal{P}))=10^{-6}$, $10^{-5}$, $10^{-4}$, and $10^{-3}$, respectively. Yellow-red color code: Compact spheres are warmer, blue: Porous grains are warmer.} 
	\label{Temperaturedistribution2_xz}
\end{figure*}
\section{Results}
\label{Section3}
In this section we present the results of the radiative transfer simulations. The goal is to understand the influence of dust grain porosity on the appearance of the disk.\\
We computed the temperature distribution in the disk for several dust masses $M_{\textrm{dust}}$. Furthermore, we calculated the spectra of scattered and reemitted light, the total intensity map, and the polarized intensity map for each dust mass and disk inclination $i \in \left\lbrace 5^\circ, 55^\circ, 80^\circ, 85^\circ \right\rbrace$. In the following, we assumed a dust grain porosity of $\mathcal{P}=0.6$. To study the influence of grain porosity, we compared the resulting disk properties and/or observational quantities with those of the corresponding reference model with compact spherical grains. Note that both disk models (with porous grains and the reference model with compact grains) contain the same total number of dust grains. Thus, the quantity $M_{\textrm{dust}}/(1-\mathcal{P})$ is fixed.\\
In the following we discuss most of our results for disks with $M_{\textrm{dust}}/(1-\mathcal{P})=10^{-5}\,\textrm{M}_\odot$ and $10^{-4}\,\textrm{M}_\odot$, which correspond to typical masses of the dust in disks around T Tauri stars (\citealt{Shu1987}; \citealt{Beckwith1990}).

\begin{figure*}[hbtp!]
    \includegraphics[height=0.45\linewidth, angle=270]{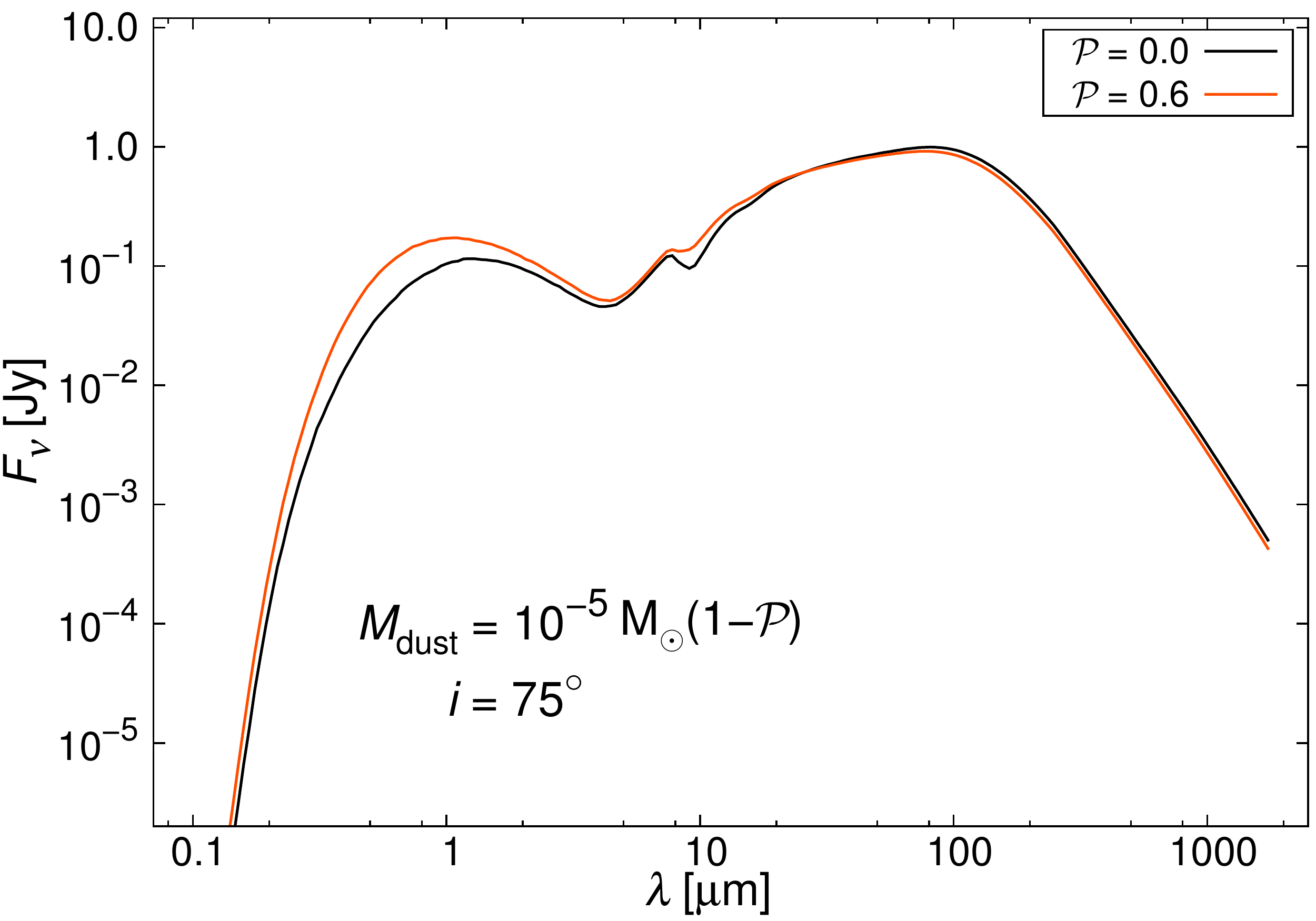}
    \includegraphics[height=0.45\linewidth, angle=270]{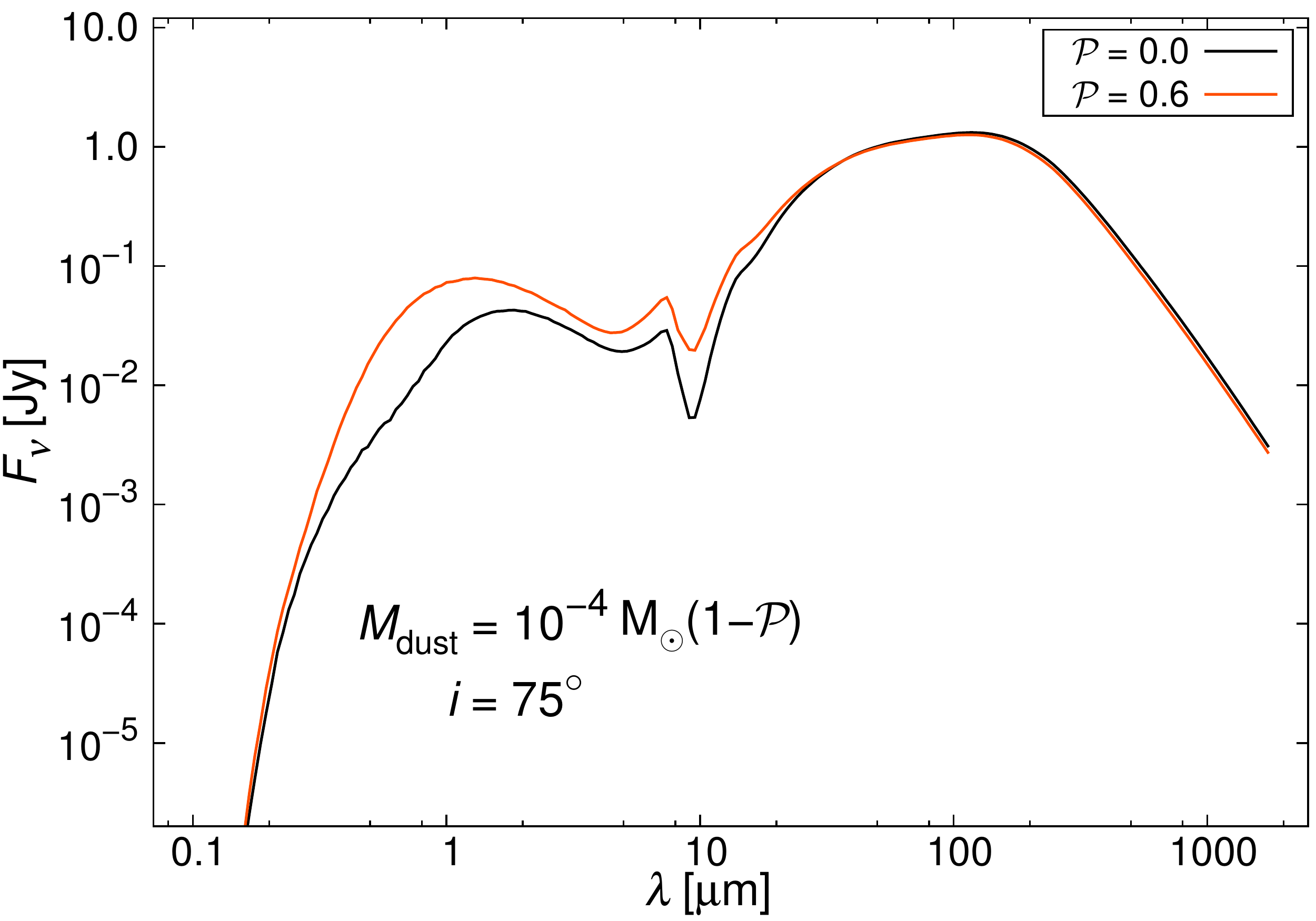}\\
    \includegraphics[height=0.45\linewidth, angle=270]{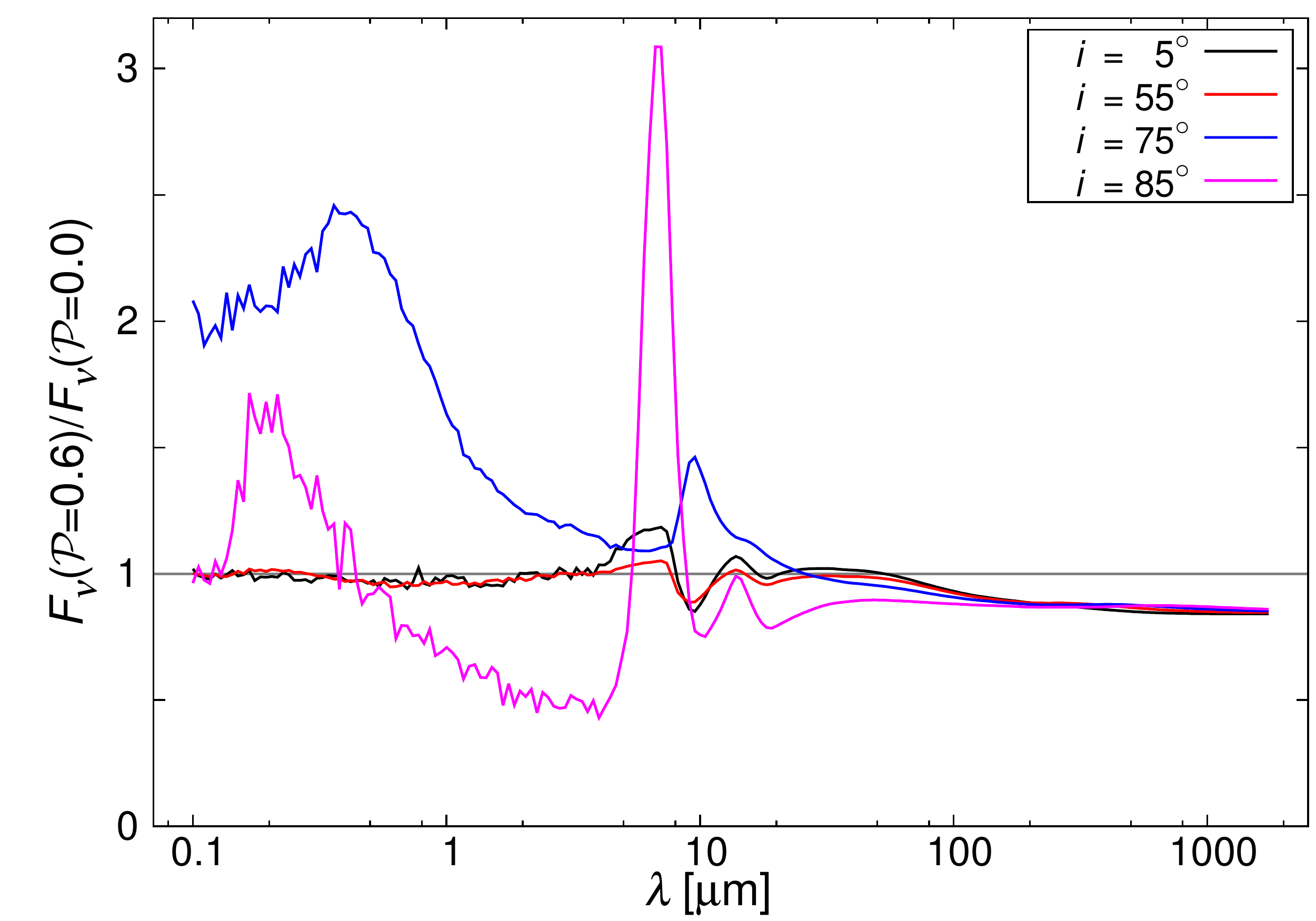}
    \includegraphics[height=0.45\linewidth, angle=270]{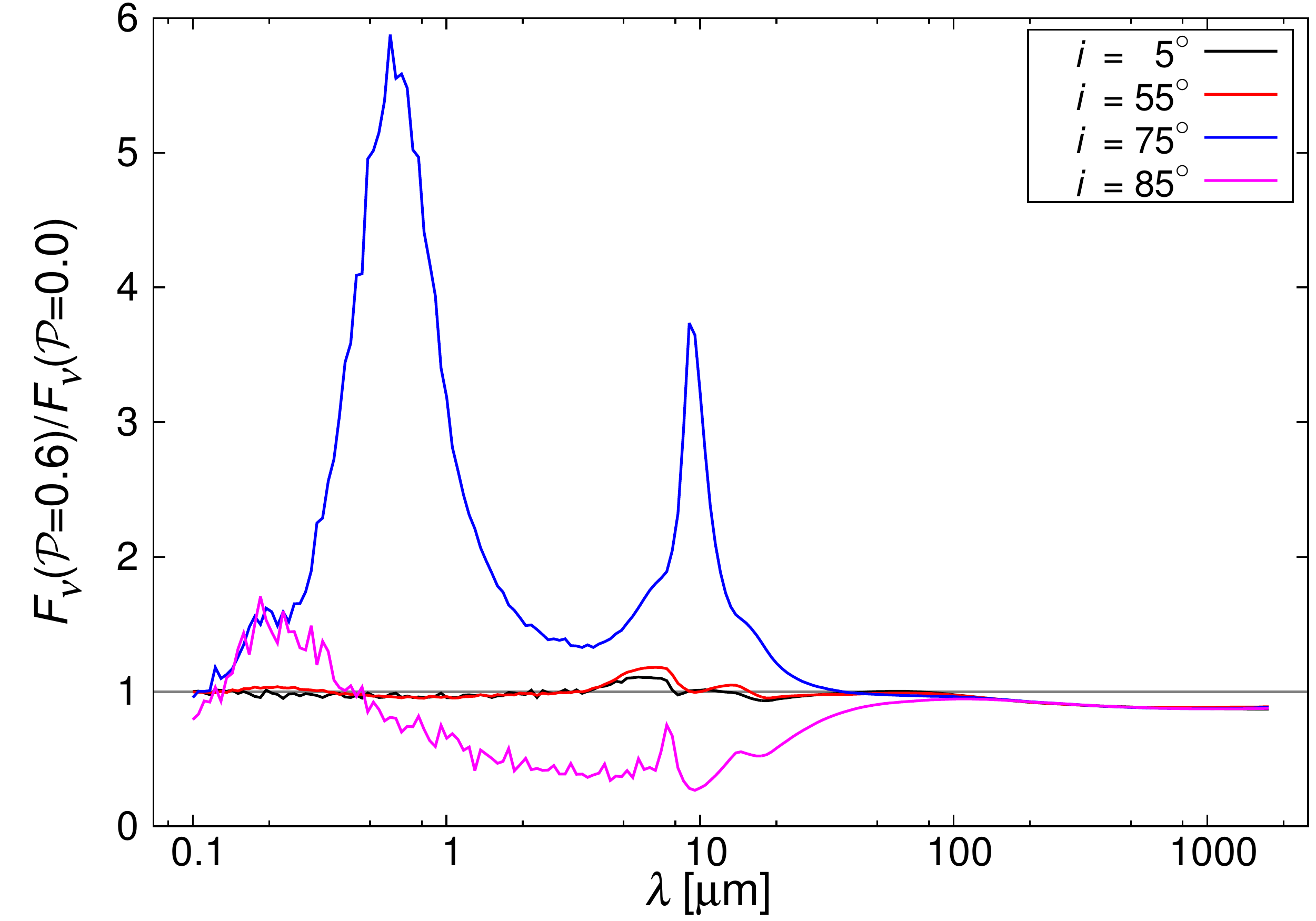}
	\caption{Spectral energy distributions for a circumstellar disk with dust masses of $M_{\textrm{dust}}=10^{-5}\,\textrm{M}_\odot\, (1-\mathcal{P})$ (left) and $M_{\textrm{dust}}=10^{-4}\,\textrm{M}_\odot\, (1-\mathcal{P})$ (right), composed of compact spherical grains ($\mathcal{P}=0.0$) or porous grains ($\mathcal{P}=0.6$). The \textit{upper} figures show the SEDs for an inclination of  $i=75^\circ$, while at the \textit{bottom} the flux ratios $F_\nu(\mathcal{P}=0.6$)/$F_\nu(\mathcal{P}=0.0$) for $i=5^\circ,\, 55^\circ,\, 75^\circ$, and $85^\circ$ are presented.\vspace*{-0.3cm}}
	\label{SEDs_M=10.0andM=100.0}
\end{figure*}

\begin{figure}[hbtp!]
\hspace*{-0.5cm}
\includegraphics[bb=142 147 453 710,clip,width=1.0\linewidth]{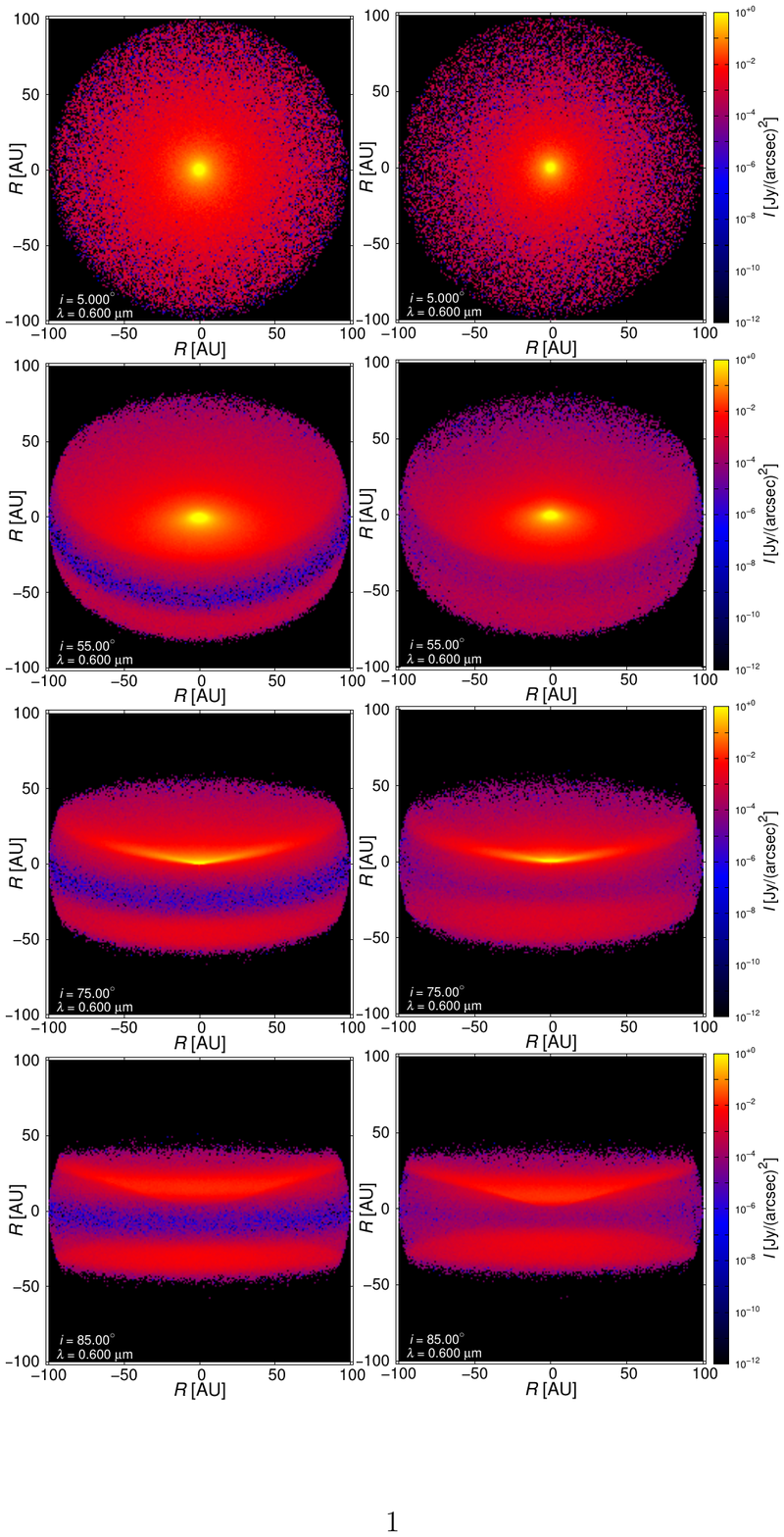}
	\caption{Scattered-light maps ($\lambda=\unit[0.6]{\mu m}$) for disks with $M_{\textrm{dust}}=10^{-5}\,\textrm{M}_\odot\, (1-\mathcal{P})$. In the left/right column scattered-light images of disks with compact spheres/porous grains ($\mathcal{P}=0.6$) are shown. Inclinations from \textit{top} to \textit{bottom}: $i =  5^\circ, 55^\circ, 75^\circ,$ and $85^\circ$.\vspace*{-0.3cm}}
	\label{Sca_Maps_M=10.0}
\end{figure}

\begin{figure}[hbtp!]
\hspace*{-0.5cm}
\includegraphics[bb=142 147 453 710,clip,width=1.0\linewidth]{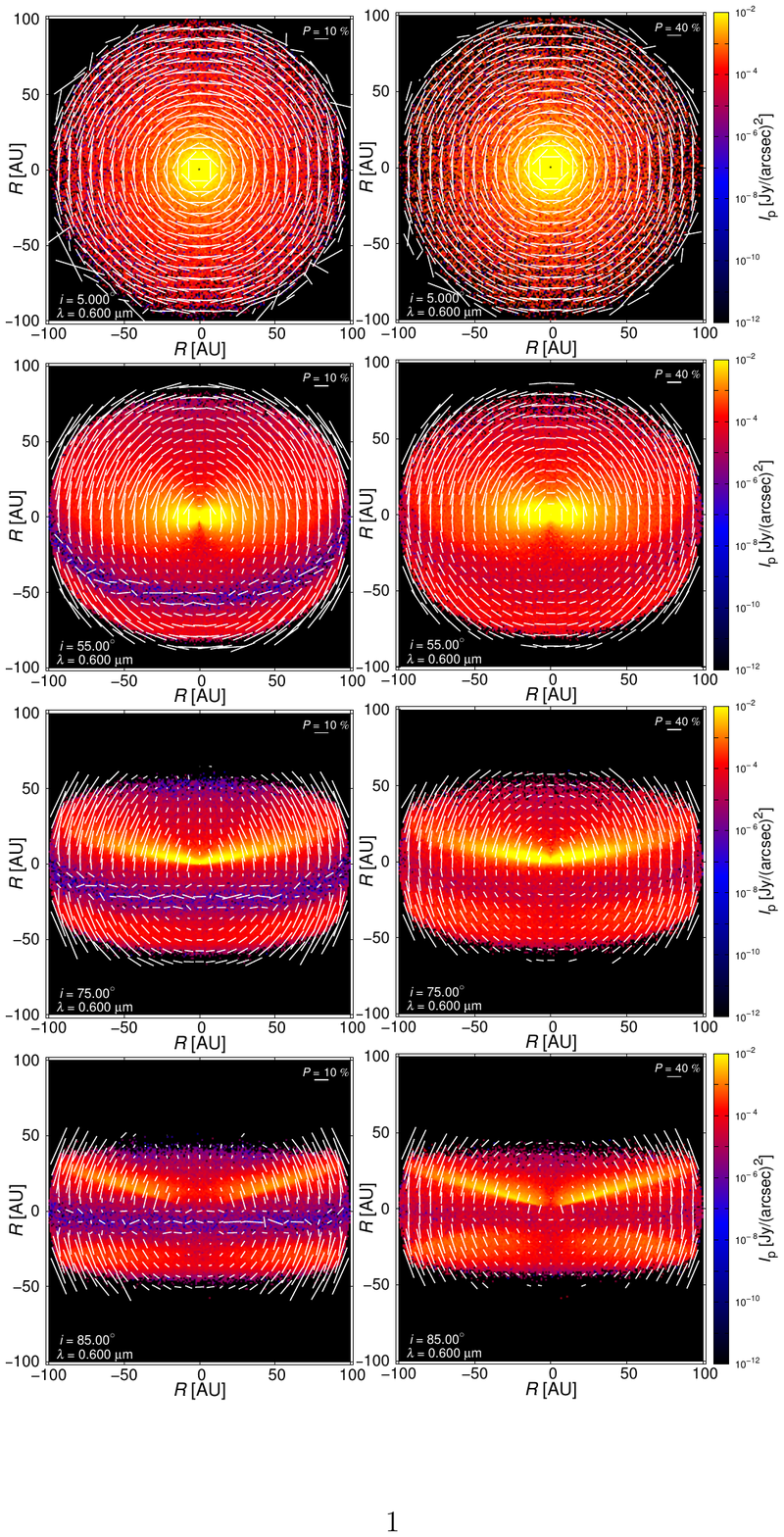}
  	\caption{Same as Fig. \ref{Sca_Maps_M=10.0} for maps of linearly polarized intensity $I_{\textrm{p}}$ and the polarization degree $P$ (white). In the left/right column polarization maps of disks with compact spheres/porous grains ($\mathcal{P}=0.6$) are shown.}
 	\label{Pol_Maps_M=10.0}
\end{figure}

\begin{figure}[hbtp!]
\hspace*{-0.5cm}
\includegraphics[bb=142 147 453 710,clip,width=1.0\linewidth]{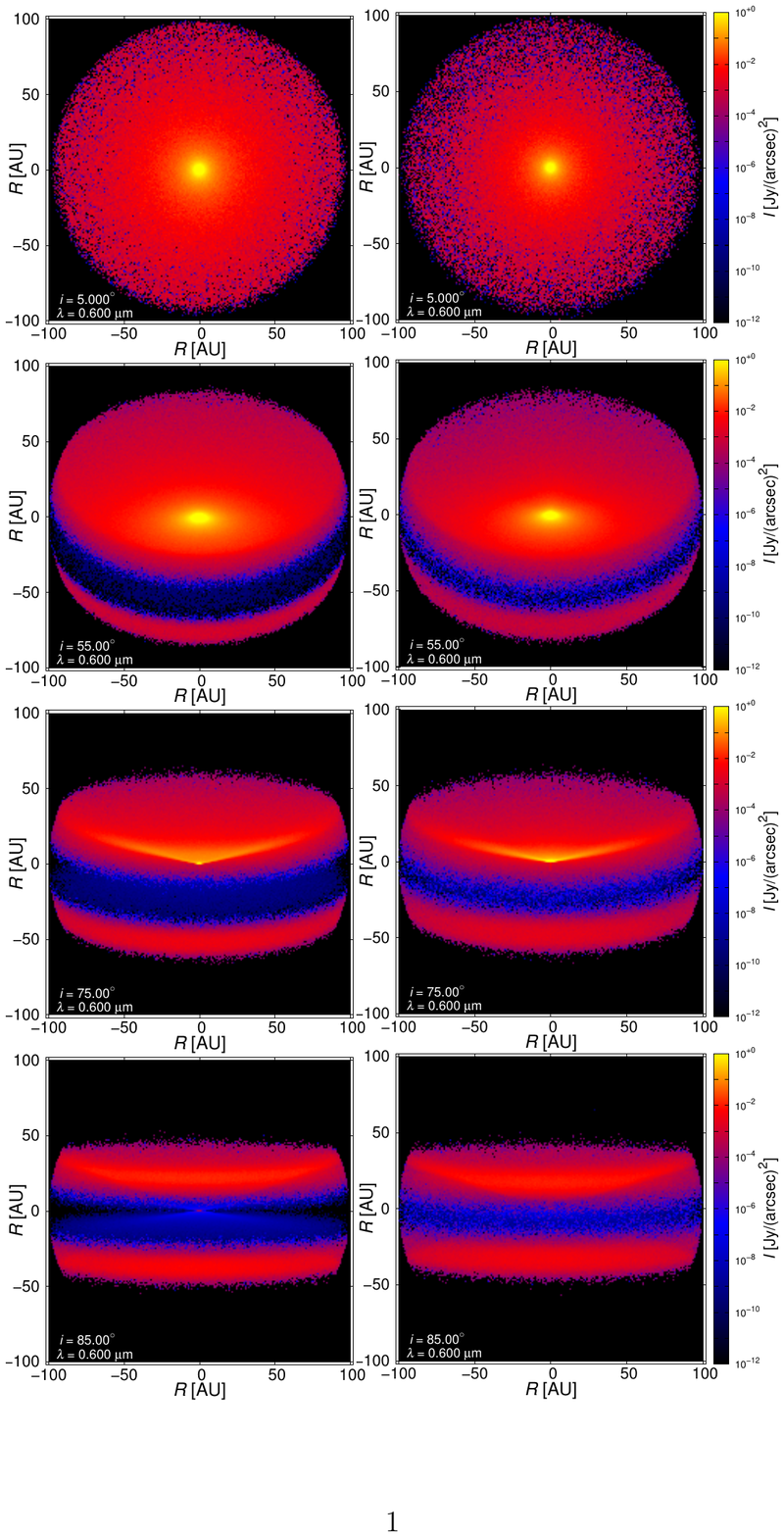}
	\caption{Same as Fig. \ref{Sca_Maps_M=10.0} for  $M_{\textrm{dust}}=10^{-4}\,\textrm{M}_\odot\, (1-\mathcal{P})$. In the left/right column scattered-light images of disks with compact spheres/porous grains ($\mathcal{P}=0.6$) are shown.}
	\label{Sca_Maps_M=100.0}
\end{figure}
\begin{figure}[hbtp!]
\hspace*{-0.2cm}
\includegraphics[height=1.0\linewidth, angle=270]{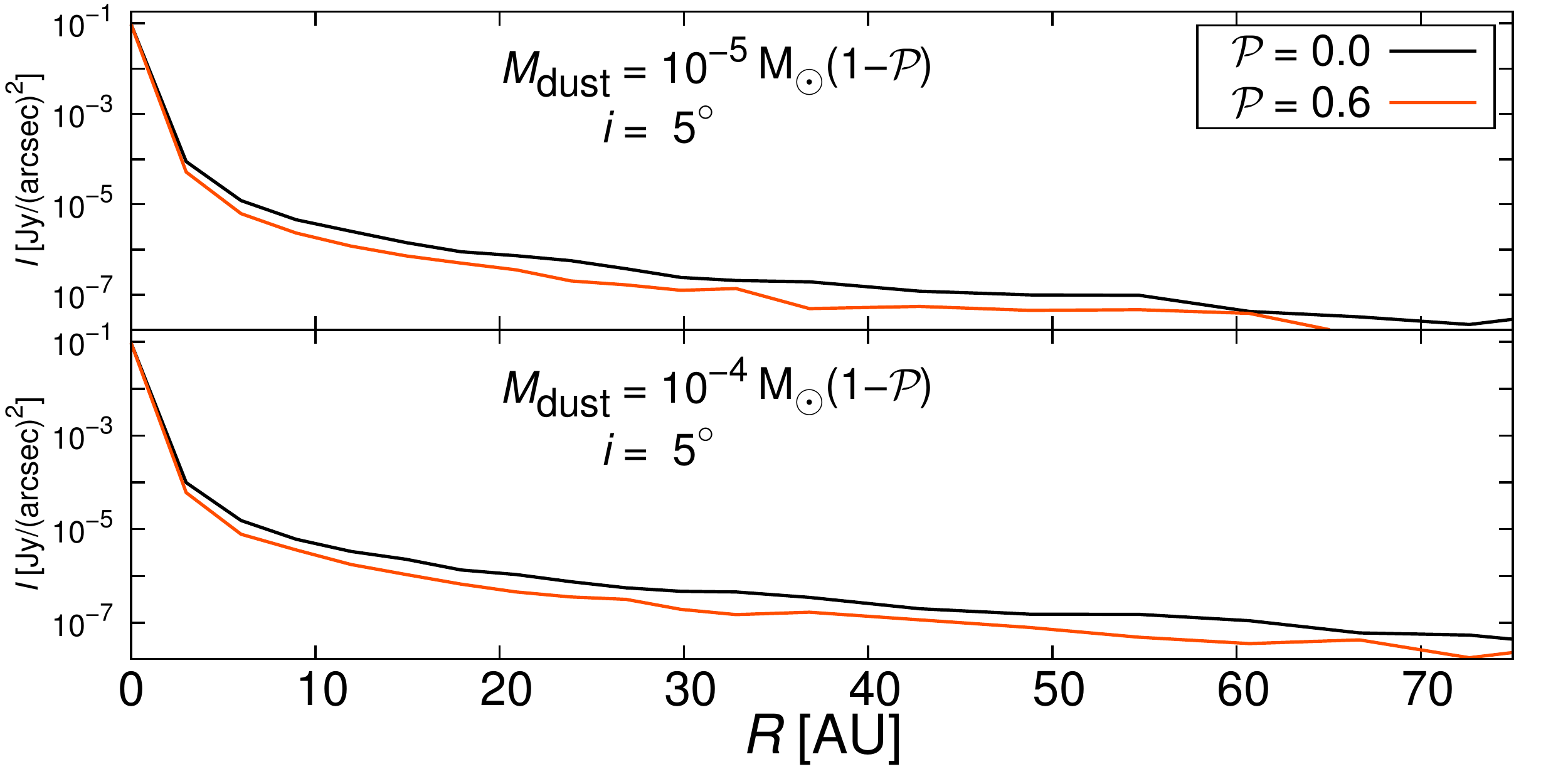}
 	\caption{Intensity of scattered light $I$ as a function of the radial distance $R$ at a disk inclination $i=5^\circ$. The intensity is slightly, but significantly higher for compact spheres.}
	\label{Flux_Quer}
\end{figure}

\begin{figure}[hbtp!]
\hspace*{-0.5cm}
\includegraphics[bb=142 147 453 710,clip,width=1.0\linewidth]{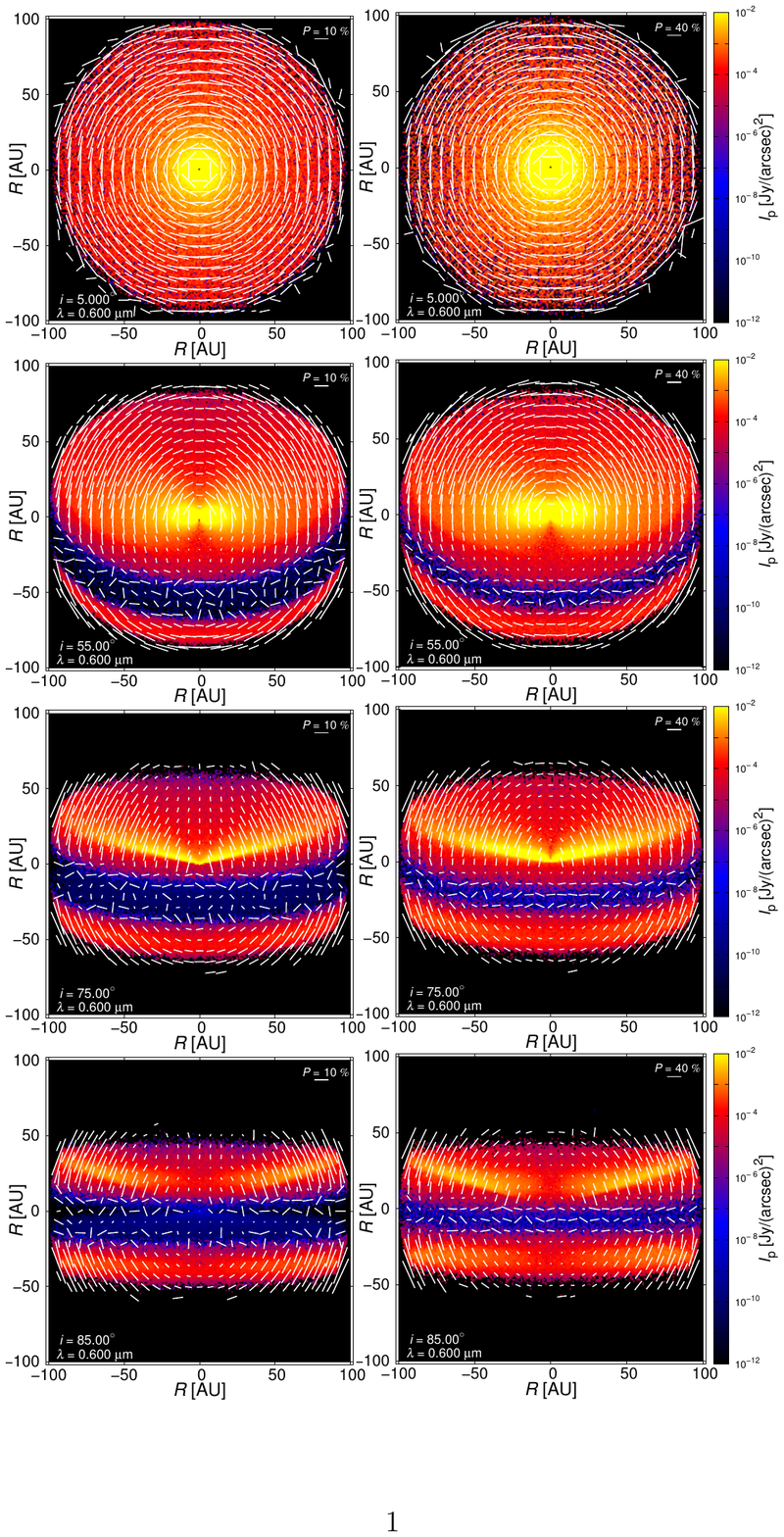}
 	\caption{Same as Fig. \ref{Sca_Maps_M=10.0} for maps of linearly polarized intensity $I_{\textrm{p}}$ and the polarization degree $P$ (white), for $M_{\textrm{dust}}=10^{-4}\,\textrm{M}_\odot\, (1-\mathcal{P})$. In the left/right column polarization maps of disks with compact spheres/porous grains ($\mathcal{P}=0.6$) are shown.}
	\label{Pol_Maps_M=100.0}
\end{figure}

\subsection{Temperature distributions}
\label{Tempdistribution}
We investigated the possible influence of porosity on the dust temperature in protoplanetary disks and calculated the temperature distribution in the plane perpendicular to the midplane to study the temperatures in intermediate regions of the disk. The dust grains in these regions are particularly interesting because on the one hand, the grains might be embedded sufficiently deep in the disk to be protected from energetic interstellar radiation. On the other hand, these layers are located significantly above the disk midplane in which gas may freeze out on dust grains. These conditions have influence on chemical processes. Complex, efficient chemical networks are expected, especially for reactions on grain surfaces, the gas-dust interaction, and the gas-phase chemical evolution (e.g., \citealt{vanDishoeck1998}; \citealt{Rawlings2001}; \citealt{kruegel2002}; \citealt{vanDishoeck2004}; \citealt{Lee2004}). To emphasize the temperature deviations between the two types of dust we computed the absolute differences of both maps, $T_{\mathcal{P}=0.0}-T_{\mathcal{P}=0.6}$ (Fig. \ref{Temperaturedistribution2_xz}).\\
We find that, apart from the dust grains in the immediate vicinity of the central star, the temperature differences between compact and porous grains are below 10 K. The temperatures of the compact spheres are higher than those of porous grains in the optically thin, upper disk regions (except for $M_{\textrm{dust}}=10^{-3}\,\textrm{M}_\odot\, (1-\mathcal{P})$) but also in selected optically thick regions close to the disk midplane. The porous grains are warmer only at the optically thin/thick transition region, in the immediate vicinity of the central star, and in the region around the disk midplane (with the exception for $M_{\textrm{dust}}=10^{-6}\,\textrm{M}_\odot\, (1-\mathcal{P})$). Increasing the dust mass results in a shift of the transition region to higher disk regions. However, the influence of different porosities on the location of this transition region is weak.


\subsection{Scattering radiation}
\label{Scatteringradiation}
Below we investigate the influence of dust porosity on scattered-light SEDs and spatially resolved maps. Within the parameter space of disk models considered we find that differences between the spectra of scattered light of the two types of dust are only small at disk inclinations $i=5^\circ$ and $55^\circ$ (Fig. \ref{SEDs_M=10.0andM=100.0}), where the deviations increase only up to 20$\,\%$ in the wavelength range $\lambda\in\left[5,20\right]\,\mu \textrm{m}$. The spectral differences between the two types become larger for increasing inclinations. At $i=75^\circ$, the flux of the scattered light is enlarged significantly for disks with porous grains (up to 2.5 times for $M_{\textrm{dust}}=10^{-5}\,\textrm{M}_\odot\, (1-\mathcal{P})$ and up to six times for $M_{\textrm{dust}}=10^{-4}\,\textrm{M}_\odot\, (1-\mathcal{P})$). For $i=85^\circ$, the maximum of the flux of the scattered light is reduced to $\sim1\,\%$ of that of face-on disks. The  flux of the scattered light of the porous grains is higher than that of compact spheres for $\lambda<\unit[0.45]{\mu m}$, and lower for $\lambda>\unit[0.45]{\mu m}$.\\
The scattered-light maps (Figs. \ref{Sca_Maps_M=10.0} and \ref{Sca_Maps_M=100.0}) at  $i=5^\circ$ appear slightly fainter for porous grains (see also Fig. \ref{Flux_Quer}). Apart from  $i=5^\circ$, all scattered-light maps show the dark lane that is characteristic for inclined disks. In particular, at inclinations $i\gtrsim 55^\circ$ the dark lane is more pronounced for compact spheres. This can be explained by the fact that the absorption cross-section of compact spheres is about one order of magnitude larger than for porous grains (Fig. \ref{opticalproperties}). Consequently, the dark lane in the scattered-light maps is wider and the flux of scattered light is reduced for compact spheres. The lower opacity of the disks with  porous grains enables one to see lower disk regions. The opacity is also the reason for the well-known decrease in the scattering maximum for increasing inclination.\\
Altogether, the behavior of the scattered-light spectra and maps depends on the opacity of the disk and thus on the dust mass $M_{\textrm{dust}}$ as well. We computed the spectra for several dust masses $M_{\textrm{dust}}$ in the range from $\unit[10^{-8}]{{M}_\odot}$ to $\unit[10^{-3}]{{M}_\odot}$. For $i=5^\circ$ and $55^\circ$ we find only small differences between the disks of the two types of dust even for much higher or much lower dust masses. At higher inclinations the deviations increase with dust mass because of the increased optical depth.


\subsection{Dust reemission}
\label{DustReemission}
As shown in previous studies (e.g., \citealt{Voshchinnikov2006}; \citealt{Kirchschlager2013}), compact spheres with radii $a\in \left[5,250\right]\,\textrm{nm}$ are in general warmer than porous grains (for porosities of up to $\sim70\%$) in the stellar radiation field. Consequently, the maximum of the dust reemission of porous grains is shifted to longer wavelengths in the optically thin case. However, this effect cannot be seen in the SEDs for the optically thick disks (Fig. \ref{SEDs_M=10.0andM=100.0}). Even more, the (sub-) millimeter flux is slightly higher for compact spheres, which might be a result of the higher opacity. The millimeter slope has a value of about $3.5$ both for compact and porous grains. This value is nearly constant for all disk inclinations and dust masses. Note that grain-growth processes can occur in evolved disks. These and the possible occurrence of carbon-rich grains might strongly influence the millimeter slope.\\
The silicate band seen in absorption or emission at $\unit[9.7]{\mu m}$ is a striking feature in the reemission spectra and becomes more pronounced for increased inclinations and increased dust masses. The absorption cross-sections at these wavelengths are larger for compact spheres than for porous grains (Fig. \ref{opticalproperties}). Therefore, the silicate feature for compact spheres is more pronounced than for porous grains.\\
Another effect arising for porous particles might be stochastic heating. The surface of a single porous grain can be considered to be composed of many small subcomponents whose mean thermal energy is similar to the energy of a single incident UV photon, resulting in time-dependent temperature fluctuations in the subcomponents (similar to PAH particles) and emission peaks particularly in the mid infrared (e.g., \citealt{Dwek1986}). Previous studies showed that the temperature of highly porous aggregates approximates the equilibrium temperature of very small particles, which constitutes a clue for the increasing significance of the subcomponents (e.g., \citealt{Greenberg1990}; \citealt{Kimura1997}).

\subsection{Polarization maps}
\label{PolarizationMaps}
In Figs. \ref{Pol_Maps_M=10.0} and \ref{Pol_Maps_M=100.0} we present the maps of the linearly polarized intensity $I_{\mathrm{p}}$ and the degree of linear polarization $P$ for the two types of dust (compact spheres and porous grains) as a function of wavelength and inclination. A striking result is the increased polarization for porous grains by a factor of about four. Compared with the scattered-light maps (Figs. \ref{Sca_Maps_M=10.0} and \ref{Sca_Maps_M=100.0}) the flux of the polarization maps is about one order of magnitude lower. The dark lane is narrower for (nearly) edge-on disks with porous grains.\\
For compact, spherical grains the polarization vectors are usually perpendicular to the plane of scattering, which is defined by the directions of the incident and scattered beam of a single scattering event. Because the grains in the optically thin, outer disk regions scatter the stellar radiation only once, they produce a centro-symmetric pattern of the polarization vector with high degrees of polarization. In the optically thick, inner regions the radiation is scattered multiple times, resulting in depolarization and in the case of (nearly) edge-on disks in an alignment of the polarization vectors parallel to the disk plane (\citealt{Bastien1988}; \citealt{Fischer1994, Fischer1996}; \citealt{Whitney1992, Whitney1993}). Because of the lower absorptivity of porous grains and the higher opacity of more massive disks, this effect is stronger for compact grains and for $M_{\textrm{dust}}=10^{-4}\,\textrm{M}_\odot\, (1-\mathcal{P})$.
\begin{figure*}[hbtp!]
\hspace*{-0.5cm}
\includegraphics[bb=122 417 432 711,clip,width=1.05\linewidth]{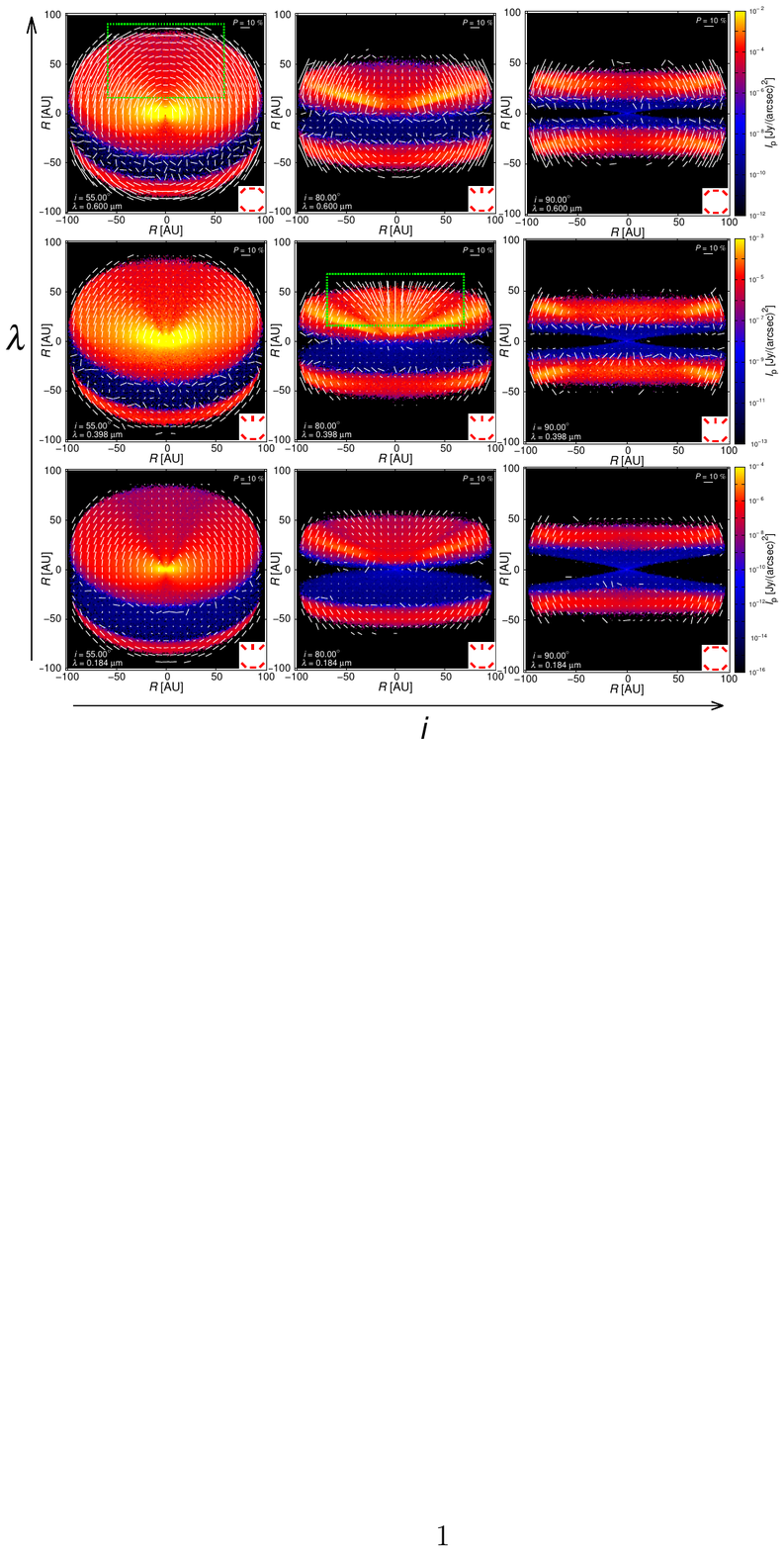}
\caption{Maps of the linear polarization $I_\textrm{p}$ for disks with $M_{\textrm{dust}}=10^{-4}\,\textrm{M}_\odot$. The wavelengths $\lambda$ and disk inclinations $i$ are chosen to illustrate the reversal of the polarization vector in selected disk regions with increasing inclination and/or wavelength. The particles are compact spheres made of astronomical silicate. In the \textit{bottom right} of each panel a sketch of the polarization pattern is plotted to distinguish whether a polarization reversal exists or not. The green box in the two single panels exemplarily indicates these cases.}
\label{9Pics}
\end{figure*}

\subsection{Polarization reversal}
\label{Sect35}
\begin{figure}[t!]
  	\centering
\includegraphics[height=1.0\linewidth, angle=-90]{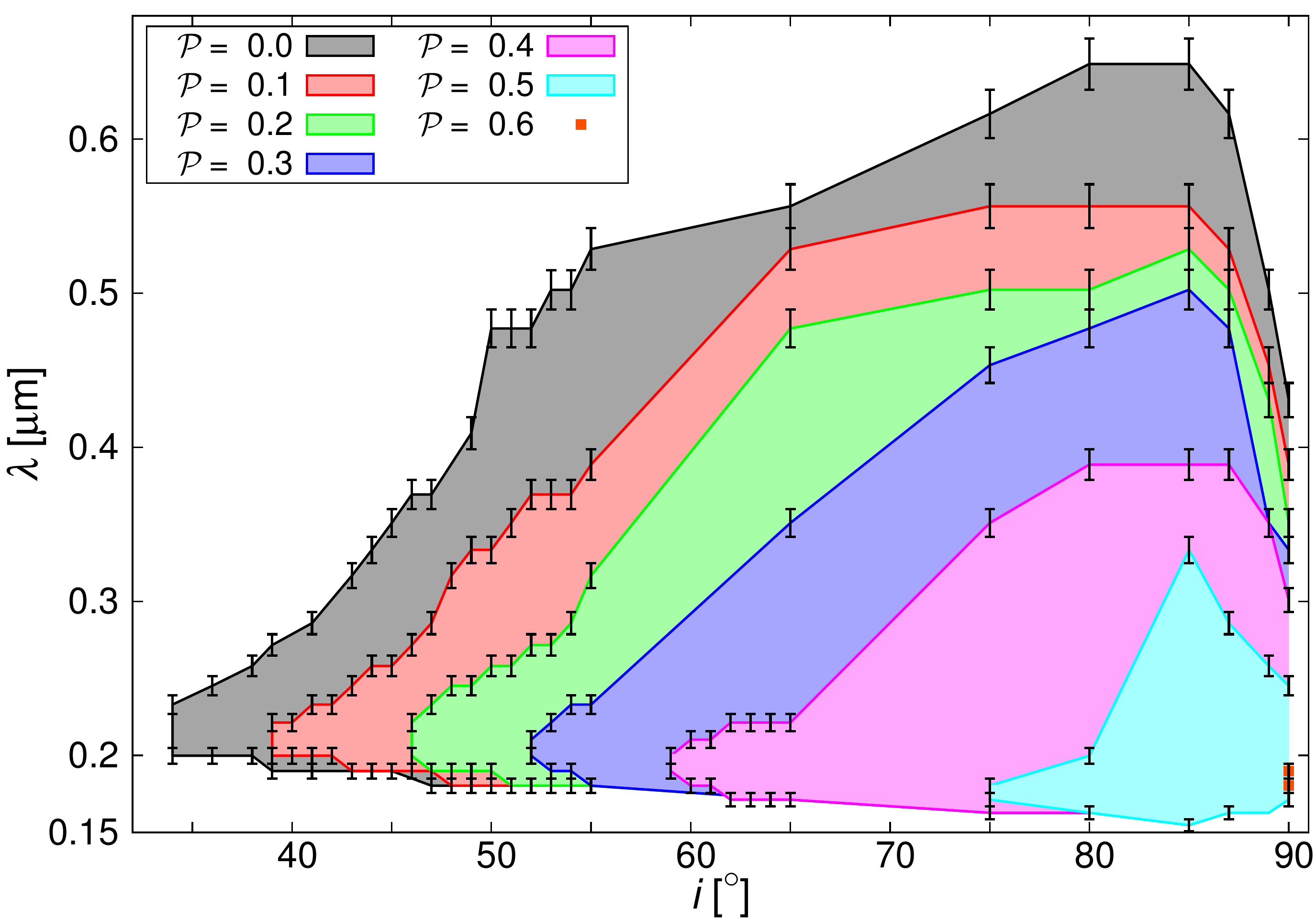}
\caption{Wavelengths at which the polarization vector in selected disk regions is radially aligned. The effect depends on the disk inclination $i$ and dust porosity $\mathcal{P}$. The effect occurs in the black colored area only for compact spheres, in the red colored area for grains with porosity $\mathcal{P}\le0.1$, etc..}
\label{Polreversal_silcate}
\end{figure}

In contrast to the polarization patterns discussed above, several scattered-light maps (compact grains, $i=75^\circ,85^\circ$) show that the polarization vectors in selected disk regions are reversed and have a  radial alignment (see Figs. \ref{Pol_Maps_M=10.0} and \ref{Pol_Maps_M=100.0}). This peculiarity was reported in previous studies. \cite{Dyck1971}, \cite{Daniel1978, Daniel1980, Daniel1982}, \cite{Simmons1982}, and \cite{Bastien1987} found a turn of the polarization direction of circumstellar dust shells, when the intensities are integrated over the whole shell. Furthermore, \cite{Fischer1994, Fischer1996} detected the appearance of radially aligned polarization vectors in selected disk regions in spatially resolved maps.\\
We begin discussing the polarization reversal for compact spheres. The polarization maps for several inclinations were calculated at a fixed wavelength ($\lambda = \unit[0.6]{\mu m}$) and a radial alignment of the polarization vector was detected in parts of the disk. The reversal occurs exclusively in the regions above the disk plane (with respect to the disk orientation in the figures). At small inclinations the polarization pattern possesses the well-known centro-symmetry (Fig. \ref{9Pics}, \textit{top left}), but with increasing inclination the polarization vector is flipped by $90^\circ$ (Fig. \ref{9Pics}, \textit{top middle}).  For even larger inclinations the polarization vector reverts (Fig. \ref{9Pics}, \textit{top right}) and the polarization pattern is centro-symmetric again.\\
Furthermore, we find that this reversal is a function of the wavelength as well (Fig. \ref{9Pics}, \textit{along the columns}).  For a wavelength of $\lambda = \unit[0.6]{\mu m}$ the polarization pattern is centro-symmetric in most parts of the disk, but with decreasing  wavelength the polarization vector is flipped by $90^\circ$. Again, the polarization vector reverts at shorter wavelengths ($\lambda < \unit[0.2]{\mu m}$).\\
\citet{Daniel1980} showed that this reversal is a function of the size parameter $x=2\pi a/\lambda$ and the dust material. As porous and nonporous grains of astronomical silicate have different scattering properties, they can be seen as consisting of different materials. This motivated us to investigate the effect of porosity on the polarization reversal. The polarization maps of disks with grains with various porosity were simulated for the following parameter space:
\begin{itemize}
\item seven grain porosities: $\mathcal{P}\in\left[0.0,0.6\right]$, in steps of $0.1$,
\item $30$ wavelengths, $\lambda \in\left[0.15,0.67\right]\,\mu\mathrm{m}$,  logarithmically  equidistantly distributed, and
\item inclinations $i \in\left[0^\circ,90^\circ\right]$.
\end{itemize}
The results for the dust mass $M_{\textrm{dust}}=10^{-4}\,\textrm{M}_\odot\, (1-\mathcal{P})$  are illustrated in Fig. \ref{Polreversal_silcate}. In the colored areas of this figure a radial alignment of the polarization vector in parts of the disk is visible. For example, at a fixed inclination of $i=55^\circ$ the reversal can be detected at wavelengths $\lambda \in\left[0.18,0.53\right]\,\mu\mathrm{m}$ for compact spheres and at wavelengths $\lambda \in\left[0.18,0.39\right]\,\mu\mathrm{m}$ for $\mathcal{P}=0.1$ grains. For $\mathcal{P}=0.6$ grains the effect appears only for $i=90^\circ$ and $\lambda \in\left[0.18,0.19\right]\,\mu\mathrm{m}$. The uncertainty in determining the reversal wavelength  is defined by the two discrete wavelengths within which the polarization vector is rotated, and marked as error bars.  For low inclinations ($i<30^\circ$) or short wavelengths ($\lambda<\unit[0.15]{\mu m}$) no reversal exists. It appears for increasing $i$ and $\lambda$ and vanishes again for long wavelengths ($\lambda> \unit[0.65]{\mu m}$). For $i=90^\circ$ the radial alignment of the polarization vector exists only for $\lambda\in\left[0.17,0.43\right]\,\mu\mathrm{m}$. Moreover, the parameter space at which the reversal appears becomes smaller when the porosity is increased.\\
We also studied the polarization reversal for $M_{\textrm{dust}}=10^{-5}\,\textrm{M}_\odot\, (1-\mathcal{P})$ with very similar quantitative results.


\subsection{Discussion of the cause of the polarization reversal}
In the discussion of the reason for the polarization reversal in several disk regions, we consider three possible explanations.
\subsubsection{Multiple scattering}
An optically thick disk  is accompanied by an anisotropic opacity distribution around the central star. \cite{Bastien1988} and \cite{Menard1988} showed that multiple scattering of unpolarized stellar light in combination with an optically thick disk and optically thin surroundings can explain the observed polarization pattern, including regions with polarization vectors parallel to the disk. \cite{Fischer1996} proposed to explain the reversal in parts of the disk by multiple scattering at the funnel-like disk surface. The photons are emitted from the central star and predominantly scattered at the $\tau=1$ surface of the flared disk, or even multiple times within the optically thick disk. To follow the direction of the observer, the photons have to be scattered again in the relatively optically thin upper layers of the disk. This multiple scattering phenomenon can create a polarization pattern that deviates in the optically thin disk parts from the common centro-symmetry and align the polarization vectors radially, explaining the results above. Furthermore, a strong dependence on the disk inclination is expected, as the considered range of scattering angles is determined by the opening angle of the $\tau=1$ surface and the disk inclination. Compared with compact spheres, porous dust grains have a reduced scattering efficiency (Fig. \ref{opticalproperties}) and an enhanced tendency for forward-scattering (\citealt{Min2012}), which both constrain the possibilities of the radial alignment of the polarization vectors, which we also observed.\\
In contrast, the wavelength dependence of the polarization alignment is not as obvious. The absorption and scattering efficiencies decrease with increasing wavelength, displacing the $\tau=1$ surface. At short wavelengths this leads to a constricted opening angle between the $\tau=1$ surface and the disk inclination, setting a lower-wavelength-limit for the radial alignment. For long wavelengths ($\lambda>\unit[0.65]{\mu\textrm{m}}$) the scattering efficiency might actually be too low to obtain substantial multiple scattering in these disk regions.\\
Thus, multiple scattering is a promising candidate for the cause of the right-angled flip of the polarization direction in selected disk regions.


\subsubsection{Particle properties}
Particle properties of the dust grains can cause polarization vectors that are not perpendicular to the scattering plane. \citet{Daniel1980} studied ellipsoidal, homogeneous envelopes and interpreted the emergent polarization reversal as an effect caused by the scattering properties of the single particles in the envelope. In analogy to the approach by \citet{Daniel1980}, we used the so-called direction parameter
\begin{align}
 \Delta=-\frac{\int_{0}^{\pi} S_{12}(\theta)\sin\theta\,\mathrm{d}\,\theta}{\int_{0}^{\pi} S_{11}(\theta)\sin\theta\,\mathrm{d}\,\theta},\label{Delta2}
\end{align}
which is a function of the optical properties of the grains. The polarization reversal occurs for wavelengths at which $\Delta$ changes its sign. We calculated the wavelength dependence of $\Delta$ for the grains considered in our study (Eq. \ref{weightedmean}) and found that  $\Delta$ is positive in the entire wavelength range. In contradiction to the simulated polarization maps, no polarization reversal would be expected.\\
The problem is the difference in the applied grain sizes. \cite{Dyck1971} proposed that the switching of the integrated polarization direction arises from the existence of different particle sizes. While \cite{Daniel1980} considered only grains of a single size and found a dependence of the reversal on the grain size parameter $x$, this simple relation cannot be adopted to our model. A weighted mean of the optical properties of a grain size distribution is used and proportions of small and larger particles contribute to the scattering properties. Moreover, if a polarization reversal existed caused only by the particle properties, the polarization vectors would be radially aligned in the entire disk and not only in parts of it. However, the results of \cite{Daniel1980} considered orientations of the net polarization integrated over the entire polarization map, and the reversal becomes noticeable only by altering the wavelength. Thus, the grain-scattering properties alone cannot explain the occurrence of the polarization reversal.


\subsubsection{The photometric and polarimetric opposition effect}
A frequently observed effect concerning the backscattering of the grains is the so called photometric and polarimetric opposition effect (e.g., \citealt{Kolokolova2011a}). This is a brightness enhancement accompanied by a polarization reversal of light scattered by comets, atmosphere-less bodies, planetary rings and interplanetary dust in the solar system and occurs for very large scattering angles only (nearly back-scattering).\\
The cause for this effect is constructive interference of electromagnetic waves scattered in particulate media (e.g., \citealt{Mishchenko2006}). It strongly depends on particle properties such as material, size, and fluffiness, as well as on the wavelength and packing density in extended media (\citealt{Kolokolova2011b}). The polarization reversal (in this context often denoted as negative polarization) for many solar system bodies and laboratory samples stretches out to $\theta\in\left[\sim150^\circ, 180^\circ \right]$. In Fig. \ref{Polarization_lambda} the polarization $P=-S_{12}/S_{11}$ is plotted as a function of the scattering angle $\theta$. Depending on the wavelength, the polarization is negative at large scattering angles. However, a brightness enhancement cannot be verified in the performed simulations, since it would be overlaid with brightness variations in the disk caused by the density and opacity distribution. Moreover, the magnitude of the negative polarization is doubtful, as the highest polarization degrees detected in the context of the polarimetric opposition effect are in the order of $\unit[1-2]{\%}$ (e.g., \citealt{Rosenbush2006}). Only in theoretical studies, such as by \cite{Mishchenko2009}, higher values were obtained.  In contrast, the polarization derived from the radiative transfer simulations can reach degrees higher than $\unit[10]{\%}$ (e.g., Fig. \ref{9Pics}, $i=80^\circ$, $\lambda=\unit[0.398]{\mu m}$).\\
Because of these deviations, there remain doubts about the accordance of the observed polarization reversal and the polarimetric opposition effect.


\subsection{Effect of polarization reversal as a diagnostic tool}
\label{ImplicationsPolreversal}
\begin{figure}[t!]
  	\centering
\includegraphics[height=1.0\linewidth, angle=-90]{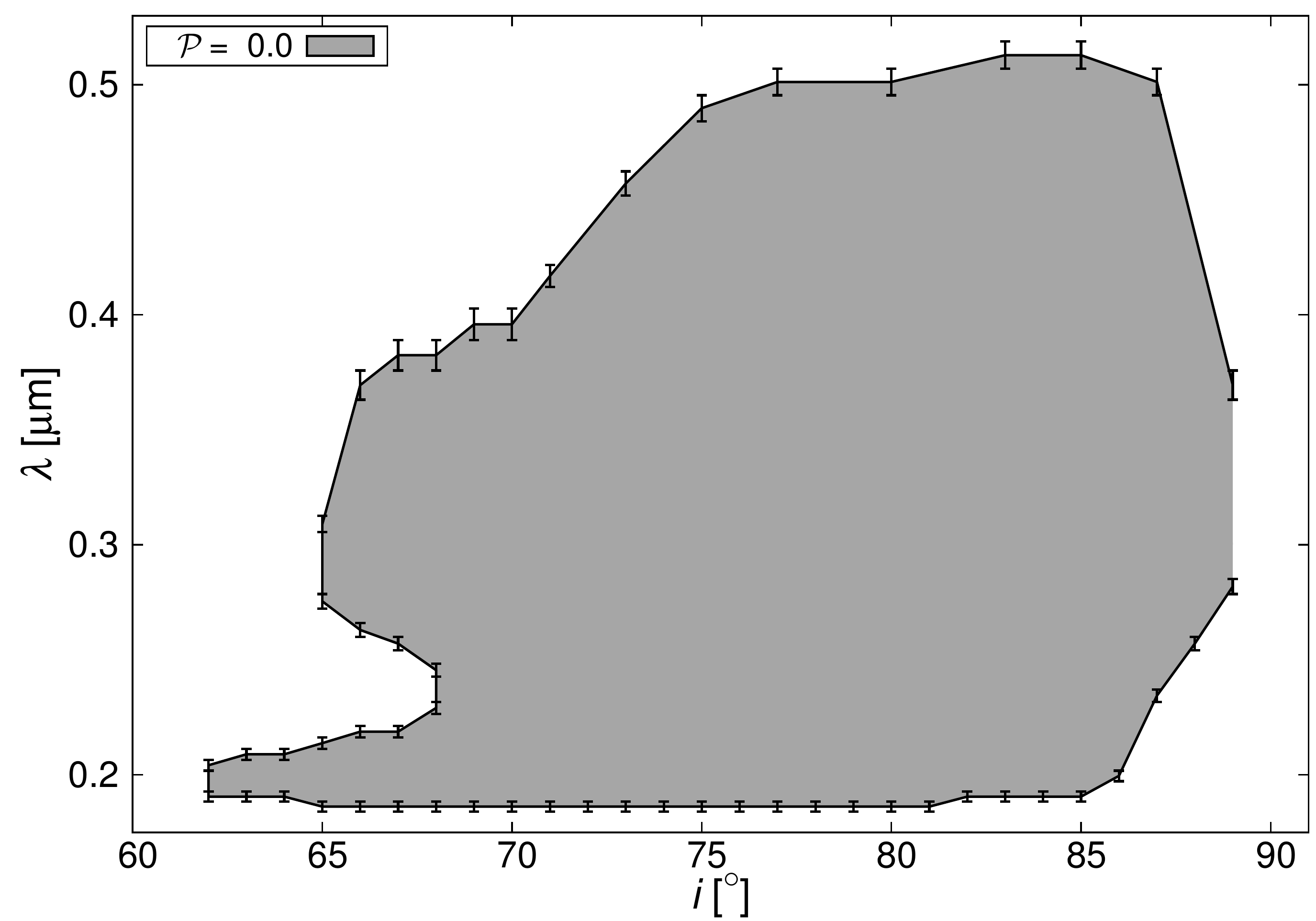}
\caption{Same as Fig. \ref{Polreversal_silcate} for compact spheres  of a mixture of $62.5\,\%$ astronomical silicate and $37.5\,\%$ graphite.}
\label{Polreversal_MRN}
\end{figure}

The polarization reversal depends on the observing wavelength, the disk inclination, and the dust properties. Thus, it provides crucial information about the dust composition. When the disk is spatially resolved and the inclination is well known, varying the observing wavelength can disclose a radial alignment of the polarization direction in selected disk regions, and the reversal wavelengths shed light on the grain properties. If the grain size distribution is further constrained (e.g., by the wavelength-dependence of the scattered light), the grain shape and in particular the grain porosity can be constrained. Thus, the polarization reversal has diagnostic potential for dust properties.\\
However, the polarization reversal is affected by several quantities. The effect is influenced by the grain size parameter and thus by the grain size distribution, which might deviate from the usual MRN distribution. In addition, the polarization reversal depends on the optical properties of the grains and thus on the chemical composition and the grain shape as well. Despite our spherical porous grain model, the existence of dust grains with a much more complex structure is conceivable, resulting in diverse polarization reversals and degeneracies. As a first outlook we investigated the polarization maps of disks with compact spheres consisting of a mixture of $62.5\,\%$ astronomical silicate and $37.5\,\%$ graphite (\citealt{WeingartnerDraine2001}) and detected a right-angled turn of the polarization vector as well (Fig. \ref{Polreversal_MRN}).\\


\section{Conclusions}
\label{Section4}
While dust grains in protoplanetary disk modeling are commonly assumed to be spherical, allowing one to apply the Mie scattering formalism, the grain shape is expected to deviate strongly from spheres in reality. In this study we particularly focused on the potential porosity of the dust and investigated its effect on various selected observable quantities of young circumstellar disks: temperature distributions, spectral energy distributions, and spatially resolved intensity and polarization maps.\\
We found that the flux in the optical wavelength range and the silicate peak at $\sim\unit[9.7]{\mu m}$ is increased for porous grains, both depending quantitatively on the disk mass. For (nearly) edge-on disks in scattering radiation the dark lane is narrower for porous grains. Compared with the scattered-light maps, the flux of the polarization maps is about one order of magnitude lower. The dark lane in the polarization maps becomes narrower for the porous grains as well. The infrared emission, and in particular the spectral index, shows only minor differences between the disks with the two types of dust. Moreover, a small but significant deviation ($<\unit[10]{K}$) exists between the dust temperatures in a disk of compact or porous grains. In particular, the porous grains are warmer than compact spheres at the optically thin/thick transition region.\\
Polarimetry is an auspicious technique for investigating the physics of dust grains in circumstellar disks. Simulated polarization maps reveal an increase in the linear polarization degree by a factor of about four for porous grains, regardless of the disk inclination. A polarization reversal is detected in selected disk regions, which depends on the observing wavelength, grain porosity, and disk inclination. We investigated the reversal wavelengths for numerous parameter combinations and exposed their interplay. Within the parameter space considered, the reversal can be detected for medium and highly inclined disks in a wavelength range between $\lambda\in\left[0.15,0.65\right]\,\mu \textrm{m}$. In addition to astronomical silicate, a dust mixture of astronomical silicate and graphite shows the effect as well. We discussed three possible explanations and found that multiple scattering seems to be the cause for the occurrence of the polarization reversal.


\begin{acknowledgements}
F.K. acknowledges financial support by the German
      \emph{Deut\-sche For\-schungs\-ge\-mein\-schaft, DFG\/}, through the project WO 857/7-1.
\end{acknowledgements}

\bibliography{aa}
\bibliographystyle{aa}

\end{document}